\documentclass[preprint,showpreprintnumbers,superscriptaddress]{revtex4}

\usepackage{graphicx}% Include figure files
\usepackage{bm}% bold math
\usepackage{amsfonts}
\usepackage{latexsym}
\usepackage{amsmath}
\usepackage[english]{babel}
\usepackage[latin1]{inputenc}
\usepackage[T1]{fontenc}
\usepackage{slashed}
\usepackage{float}
\usepackage{subfigure}

  \newcommand{\be}{\begin{equation}}
  \newcommand{\ee}{\end{equation}}
  \newcommand{\ba}{\begin{eqnarray}}
  \newcommand{\ea}{\end{eqnarray}}
  \newcommand{\eite}{\end{itemize}}
  \newcommand{\bite}{\begin{itemize}}
  \newcommand{\betabeta}{\mbox{$(\beta \beta)_{0 \nu}  $}}

\def\ltap{\ \raisebox{-.4ex}{\rlap{$\sim$}} \raisebox{.4ex}{$<$}\ }
\def\gtap{\ \raisebox{-.4ex}{\rlap{$\sim$}} \raisebox{.4ex}{$>$}\ }

 \newcommand{\meff}{\mbox{$\left|  < \!  m \!  > \right| \ $}}
 
 \newcommand{\nn}{\nonumber}

\begin{document}
 
\preprint{COLO-HEP-564, SISSA 43/2011/PE, UCI-TR-2011-19}

\title{Predictions for Neutrino Masses, $\beta\beta_{0\nu}$-Decay 
and Lepton Flavor Violation in a SUSY $SU(5) \times T^{\prime}$ 
Model of Flavour}

\author{Mu-Chun Chen}
\email[]{muchunc@uci.edu}
\affiliation{Department of Physics \& Astronomy, University of California, Irvine, CA 92697-4575, U.S.A.}

\author{Kalyana T. Mahanthappa}
\email[]{ktm@colorado.edu}
\affiliation{Department of Physics, University of Colorado, 
Boulder, CO 80309-0390, U.S.A.}

\author{Aurora Meroni}
\email[]{ameroni@sissa.it}
\affiliation{SISSA and INFN-sezione di Trieste, 
Via Bonomea 265, 34136 Trieste, Italy}
% \affiliation{Instituto Nazionale di Fisica Nucleare, 
% Sezione di Trieste, Via Valerio 2, 34126 Trieste, Italy}

\author{S. T. Petcov
\footnote{Also at: Institute of Nuclear Research and
Nuclear Energy, Bulgarian Academy of Sciences, 
1784 Sofia, Bulgaria}
}
% \email[]{petcov@sissa.it}
\affiliation{SISSA and INFN-sezione di Trieste, Via Bonomea 265, 
34136 Trieste, Italy}
% \affiliation{Instituto Nazionale di Fisica Nucleare, Sezione 
% di Trieste, Via Valerio 2, 34126 Trieste, Italy}
\affiliation{IPMU, University of Tokyo, Tokyo, Japan}

% \date{today}

\begin{abstract}
We obtain predictions for the neutrino masses, 
the effective Majorana mass in neutrinoless double beta 
decay and for the rates of the lepton flavor violating processes 
$\mu\rightarrow e + \gamma$, $\tau \rightarrow e + \gamma$
and  $ \tau \rightarrow \mu + \gamma$
in a SUSY $SU(5) \times T^{\prime}$ Model of flavour, 
which gives rise to realistic masses and mixing 
patterns for quarks and leptons.
\end{abstract}

%\begin{keyword}


%\end{keyword}

\maketitle

\section{Introduction}

 Understanding the origin of the patterns of neutrino 
masses and mixing, emerging from the neutrino oscillation, 
$^3H$ $\beta-$decay, etc. data is one of the most 
challenging problems in neutrino physics. 
It is part of the more general fundamental problem 
in particle physics of understanding the origins of 
flavour, i.e., of the patterns 
of the quark, charged lepton and neutrino masses 
and of the quark and lepton mixing.

   At present we have compelling evidence for 
existence of mixing of three light massive neutrinos 
$\nu_i$, $i=1,2,3$, in the weak charged lepton current 
(see, e.g., \cite{PDG10}). The masses $m_i$ of the three  
light neutrinos $\nu_i$ do not exceed approximately 
1 eV, $m_i\ltap 1$ eV, i.e., they are much smaller than the 
masses of the charged leptons and quarks.
The three light neutrino mixing is described 
(to a good approximation)
by the Pontecorvo, Maki, Nakagawa, Sakata (PMNS) $3\times 3$ 
unitary mixing matrix, $U_{\rm PMNS}$. In the widely used 
standard parametrisation \cite{PDG10},  $U_{\rm PMNS}$ is 
expressed in terms of the solar, atmospheric and reactor 
neutrino mixing angles $\theta_{12}$,  $\theta_{23}$ and 
$\theta_{13}$, respectively, and one Dirac - $\delta$, and 
two Majorana \cite{BHP80} - $\alpha_{21}$ and $\alpha_{31}$, 
CP violating phases:
%%%%%%%%%%%%%%%%%%%%%%%%%%%
\be
 U_{PMNS} \equiv U = V(\theta_{12},\theta_{23},\theta_{13},\delta)\,
Q(\alpha_{21},\alpha_{31})\,,
\label{UPMNS}
\ee
%%%%%%%%%%%%%%%%%%%%%%%%%%%
%
where
%%%%%%%%%%%%%%%%%%%%%%%%%%%%%%%
\be % U_{PMNS}=
V = \left(
     \begin{array}{ccc}
       1 & 0 & 0 \\
       0 & c_{23} & s_{23} \\
       0 & -s_{23} & c_{23} \\
     \end{array}
   \right)\left(
            \begin{array}{ccc}
              c_{13} & 0 & s_{13}e^{-i\delta} \\
              0 & 1 & 0 \\
              -s_{13}e^{i\delta} & 0 & c_{13} \\
            \end{array}
          \right)\left(
                   \begin{array}{ccc}
                     c_{12} & s_{12} & 0 \\
                     -s_{12} & c_{12} & 0 \\
                     0 & 0 & 1 \\
                   \end{array}
                 \right)\,, 
% diag(1, e^{i\alpha_{21}/2},e^{i\alpha_{31}/2})\,,
\label{V}
\ee
%%%%%%%%%%%%%%%%%%%%%%%%%%%%%%%%%%%
%
and we have used the standard notation 
$c_{ij} \equiv \cos\theta_{ij}$,
$s_{ij} \equiv \sin\theta_{ij}$, and
%%%%%%%%%%%%%%%%%%%%%%%%%%%%%%%%%
\be
Q =  diag(1, e^{i\alpha_{21}/2},e^{i\alpha_{31}/2})\,.
\label{Q}
\ee
%%%%%%%%%%%%%%%%%%%%%%%%%%%%%%%
%
  
  It is known  since rather long time 
from the analysis of the  neutrino oscillation 
data that $\theta_{23}$ and $\theta_{12}$  
are ``large'' while $\theta_{13}$ is ``small'' 
(see, e.g., \cite{PDG10}):
$\theta_{23} \cong \pi/4$, 
$\theta_{12} \cong \sin^{-1}(\sqrt{0.3})$ and 
 $\sin^2\theta_{13} \ltap 0.05$ (at $3\sigma$).
Recently the T2K collaboration reported 
\cite{Abe:2011sj} evidence at $2.5\sigma$ 
for a non-zero value of the ``reactor''
angle $\theta_{13}$.
Subsequently the MINOS collaboration also reported   
evidence for a 
nonzero value of $\theta_{13}$, 
although with a smaller statistical 
significance \cite{MINOS240611}. A global analysis of 
the neutrino oscillation data, including 
the data from the T2K and MINOS experiments, performed 
in \cite{Fogli:2011qn} showed  that actually 
$\sin\theta_{13}\neq 0$ at $\geq 3\sigma$.
The results of this analysis, in which the 
neutrino mass squared differences 
$\Delta m^2_{21} \equiv \Delta m^2_{\odot}$ 
\newpage
%%%%%%%%%%%%%%%%%%%%%%%%%%%%%%%%%%%%%%%%
\begin{table}[b!]
\centering \caption{
\label{tab:tabdata-1106} The best-fit values and
$3\sigma$ allowed ranges of the 3-neutrino oscillation parameters,
obtained from a global fit of the current neutrino oscillation data, 
including the recent T2K and MINOS results (from ~\cite{Fogli:2011qn}).}
\renewcommand{\arraystretch}{1.1}
\begin{tabular}{lcc}
\hline \hline
 Parameter  &  best-fit ($\pm 1\sigma$) & 3$\sigma$ \\ \hline
 $\Delta m^{2}_{\odot} \; [10^{-5}eV^2]$   & 7.58$^{+0.22}_{-0.26}$ & 
               6.99 - 8.18 \\
$ |\Delta m^{2}_{A}| \; [10^{-3}eV^2]$ & 2.35$^{+0.12}_{-0.09}$  & 
           2.06 - 2.67\\
 $\sin^2\theta_{12}$  & 0.306$^{+0.018}_{-0.015}$ 
            & 0.259 - 0.359\\
 $\sin^2\theta_{23}$  & 0.42$^{+0.08}_{-0.03}$ &  0.34 - 0.64 \\
$\sin^2\theta_{13}$   &  0.021$^{+0.07}_{-0.08}$  &  0.001-0.044 \\
\hline\hline
\end{tabular}
\end{table}
%%%%%%%%%%%%%%%%%%%%%%%%%%%%%%%%%%%%%%%%
%
\noindent and $\Delta m^2_{31} \equiv \Delta m^2_{\rm A}$,
responsible for the solar, and the dominant atmospheric,
neutrino oscillations, were determined as well, are shown 
in Table \ref{tab:tabdata-1106}. 
The authors of \cite{Fogli:2011qn} find, in particular, 
the following  best fit value and  $3\sigma$
allowed range of  $\sin^2\theta_{13}$:
%%%%%%%%%%%%%%%%%%%%%%%%%
\begin{equation}
 \sin^2\theta_{13} = 0.021\,,~~~~
% \sin^2\theta_{13} = 0.021~(0.025) \pm 0.007\,,
% 0.001~(0.005) \leq \sin^2\theta_{13} \leq 0.044~(0.050\,,
0.001 \leq \sin^2\theta_{13} \leq 0.044\,,
\label{th13Fogli}
\end{equation}
%%%%%%%%%%%%%%%%%%%%%%%%
%
using the ``old'' fluxes of reactor $\bar{\nu}_e$ in the analysis 
(see \cite{Fogli:2011qn} for further details). 
Moreover, it was found in the same global 
analysis that $\cos\delta = -1$ 
(and $\sin\theta_{13}\cos\delta = - 0.14$) 
is clearly favored by the data over 
$\cos\delta = + 1$ 
(and $\sin\theta_{13}\cos\delta = + 0.14$)~
\cite{footn1}.
It is interesting to note that 
the best fit value of the Dirac 
CP violating phase $\delta$,
obtained in the analysis of the atmospheric neutrino data
by the Super Kamiokande (SuperK) Collaboration reported 
in~\cite{superK2010}, reads  
$\delta = 220^{\circ}\cong 1.22\pi$.
A not very different best fit value of $\delta$ was found
in the very recent global neutrino oscillation data 
analysis performed in \cite{SchwT2V0811}:
$\delta = -0.61\pi~(-0.41\pi)$, or equivalently 
$\delta = 1.39\pi~(1.59\pi)$, the value quoted 
corresponding to neutrino mass spectrum with normal 
(inverted) ordering. It should be added, however, 
that, except possibly for the negative 
sign of $\cos\delta$, the preference of the data 
for the indicated specific values of 
the phase $\delta$ are at present 
statistically insignificant.

  The T2K and MINOS results will be tested in the upcoming 
reactor neutrino experiments Double Chooz \cite{DCHOOZ}, 
Daya Bay \cite{DayaB} and RENO \cite{RENO}.
If confirmed, they will have far reaching implications 
for the program of future research in neutrino physics 
(see, e.g., \cite{MMexTSchwth13}). 

 The experimentally determined  values of 
the solar and atmospheric neutrino mixing angles 
$\theta_{12}$ and $\theta_{23}$ coincide, or are close, 
to those predicted in the case of
tri-bimaximal mixing (TBM)~\cite{tri1}:
%%%%%%%%%%%%%%%%%%%%%%%%%%%
\begin{equation}
\sin^{2} \theta_{23}^{\mbox{\tiny TBM}} = 1/2 \; , 
\quad \sin^{2}\theta_{12}^{\mbox{\tiny TBM}} = 1/3 \; , 
\quad \sin\theta_{13}^{\mbox{\tiny TBM}} = 0 \;,
\label{TBMth1}
\end{equation}
%%%%%%%%%%%%%%%%%%%%%%%%%
%
with the TBM mixing matrix given by:
%%%%%%%%%%%%%%%%%%%%%%%%%%%%%%%%%%%%%%%%%%
\begin{equation}
U_{TBM} = \left(\begin{array}{ccc}
\sqrt{2/3} & \sqrt{1/3} & 0 \\
-\sqrt{1/6} & \sqrt{1/3} & -\sqrt{1/2} \\
-\sqrt{1/6} & \sqrt{1/3} & \sqrt{1/2}
\end{array}\right) \;.
\label{TBMM}
\end{equation}
%%%%%%%%%%%%%%%%%%%%%%%%%%%%%%%%
%

It is appealing to assume, following 
\cite{tri1}, that in leading approximation 
we have (up to diagonal phase matrices)
$U_{\rm PMNS}\cong U_{\rm TBM}$, the reason being 
that the specific form of $U_{\rm TBM}$ can be 
understood 
on the basis of symmetry considerations.
However, in order to account for the 
current neutrino mixing data, and more 
specifically, for the fact 
that $\theta_{13} \neq 0$, 
$U_{TBM}$ has to be ``corrected''. 
Such a correction can naturally arise given 
the fact that, as is well known, 
the PMNS matrix receives, in general, 
contributions from the diagonalisation of the 
neutrino and charged lepton mass matrices, $U_\nu$ 
and  $U_\ell^\dagger$, respectively:
%%%%%%%%%%%%%%%%%%%%%%
\be 
\label{UeUnu1}
% U_{\rm PMNS} 
U = U_\ell^\dagger \, U_\nu~ = 
\tilde{U}_\ell^\dagger\, \tilde{\Phi} \tilde{U}_\nu\,Q\,,
\ee
%%%%%%%%%%%%%%%%%%%
%
where \cite{FPR} $\tilde{U}_\ell^\dagger$ 
and  $\tilde{U}_\nu$ are $V$-like $3\times 3$   
unitary matrices (see eq. (\ref{V})) each containing, 
in general, one Dirac-like CP violation phase,  
$\tilde{\Phi}$ is a diagonal phase matrix containing, in general,
two CP violation phases, and $Q$ is define in eq. (\ref{Q}).
In the   SUSY $SU(5) \times T^{\prime}$ model of flavour 
we are going to consider in what follows 
\cite{Chen:2007afa,Chen:2009gf}, 
$\tilde{U}_\nu$ coincides with $U_{TBM}$, 
$\tilde{U}_\nu = U_{TBM}$, and the requisite 
``correction'' leading to $\theta_{13}\neq 0$ 
is provided by $\tilde{U}_\ell^\dagger$.
This possibility was investigated 
phenomenologically by many authors, see, e.g., 
\cite{tri2,HPR07,Marzocca:2011dh} as well as 
\cite{FPR,andrea}. 
 
  It has been realized a rather long time ago that the 
TBM matrix can arise from an underlying $A_{4}$ flavour 
symmetry~\cite{Ma:2001dn}. 
Nevertheless, generating the quark mixing 
utilizing the $A_{4}$ symmetry encounters a number of 
difficulties ~\cite{Ma:2006sk}. As a consequence, 
the incorporation of the $A_4$ symmetry in GUTs 
is not straightforward,
being plagued with complications.
On the other hand, the group 
$T^{\prime}$~\cite{Chen:2007afa,frampton}, which is 
the double covering of $A_{4}$, can successfully account 
for the quark masses and mixing.
In~\cite{Chen:2007afa,Chen:2009gf} a viable SUSY model of 
flavour, including the CP violation in the quark sector,
based on the $SU(5)\times T'$ symmetry was constructed. 
The model includes three right-handed neutrino 
fields which possess a Majorana 
mass term. The light neutrino masses 
are generated by the type I see-saw mechanism and are 
naturally small. The light and heavy neutrinos are 
Majorana particles. The model is free of 
discrete gauge anomalies ~\cite{Luhn:2008xh,Chen:2006hn}.
In addition to giving rise to realistic masses 
and mixing patterns for the leptons and quarks, 
the  $SU(5)\times T'$ model under discussion 
exhibits a number of unique features. In particular,  the CP violation, 
predicted by the model, can entirely be geometrical 
in origin ~\cite{Chen:2009gf}. This interesting 
aspect of the $SU(5)\times T'$ model we will consider 
is a consequence of one of the 
special properties of the group $T'$, namely, 
that its group theoretical Clebsch-Gordon 
(CG) coefficients are intrinsically complex~\cite{cg}. 
More specifically, the only source of CP violation, e.g., 
in the lepton sector of the model is the Dirac phase $\delta$.
The Majorana phases $\alpha_{21}$ and 
$\alpha_{31}$ are predicted (to leading order) 
to have CP conserving values. 
The Dirac phase is induced effectively by the complex CG 
coefficients of the group $T'$.
The model allows
for a successful leptogenesis, 
the Dirac phase $\delta$ providing the requisite CP 
violation for the generation of the observed baryon 
asymmetry of the Universe ~\cite{Chen:2011tj}. 
As a consequence, there is a strong connection 
between the CP violation in neutrino oscillations 
and the matter-antimatter asymmetry of the Universe.

  In the present article we  investigate some aspects of the 
low energy phenomenology  of the  $SU(5)\times T'$ model 
of flavour~\cite{Chen:2007afa,Chen:2009gf}, which allows to describe 
in a unified way the masses and the mixing of  
both the quarks and the leptons, neutrinos included.
We concentrate first on the predictions of the model for the 
absolute scale of neutrino masses, the neutrino mass spectrum 
and the effective Majorana mass in neutrinoless double beta
($\betabeta$-) decay. We derive next detailed predictions 
for the rates of the lepton flavour violating (LFV) 
charged lepton radiative decays $\mu \rightarrow e + \gamma$, 
$\tau \rightarrow e + \gamma$ and $\tau \rightarrow \mu + \gamma$.

  The paper is organized as follows.
Section 2 is devoted to a rather detailed description of the 
$SU(5)\times T^{\prime}$ model of interest.
In Section 3  the predictions of the Model
for the neutrino masses, the type of neutrino mass spectrum 
and the $\betabeta$-decay effective Majorana mass are 
obtained. In Section 4 we present results on  
the rates of the LFV charged lepton radiative decays 
$\mu \rightarrow e + \gamma$, 
$\tau \rightarrow e + \gamma$ and 
$\tau \rightarrow \mu + \gamma$, calculated 
assuming the mSUGRA scenario of soft SUSY breaking.
Section 5 contains a Summary of the results obtained 
in the present work.

%%%%%%%%%%%%%%%%%%%%%%%%%%
%
\section{The $SU(5)\times T^{\prime}$ Model}
%
%%%%%%%%%%%%%%%%%%%%%%%%

The content of the chiral superfields of the model considered 
(including the three generations of matter fields, 
the $SU(5)$ Higgses in the Yukawa sector, and flavon fields) 
as well as their quantum numbers with respect 
to $SU(5)$, $T^{\prime}$, and $Z_{12} \times Z^\prime_{12}$  
symmetry groups are given in Table~\ref{tbl:charge}.  
%%%%%%%%%%%%%%%%%%%%%%%%%%
\begin{table}[t!]
\begin{tabular}{|c|cccc|ccc|cccccc|ccc|}\hline
& $T_{3}$ & $T_{a}$ & $\overline{F}$ & $N_{lR}$ & $H_{5}$ & $H_{\overline{5}}^{\prime}$ & $\Delta_{45}$ & $\phi$ & $\phi^{\prime}$ & $\psi$ & $\psi^{\prime}$ & $\zeta$ & $\zeta^{\prime} $ & $\xi$ & $\eta$  & $S$ \\ [0.3em] \hline\hline
SU(5) & 10 & 10 & $\overline{5}$ & 1 &  5 & $\overline{5}$ & 45 & 1 & 1 & 1 & 1& 1 & 1 & 1 & 1 & 1\\ \hline
$T^{\prime}$ & 1 & $2$ & 3 & 3 & 1 & 1 & $1^{\prime}$ & 3 & 3 & $2^{\prime}$ & $2$ & $1^{\prime\prime}$ & $1^{\prime}$ & 3 & 1 & 1 \\ [0.2em] \hline
$Z_{12}$ & $\omega^{5}$ & $\omega^{2}$ & $\omega^{5}$ & $\omega^{7}$ & $\omega^{2}$ & $\omega^{2}$ & $\omega^{5}$ & $\omega^{3}$ & $\omega^{2}$ & $\omega^{6}$ & $\omega^{9}$ & $\omega^{9}$ 
& $\omega^{3}$ & $\omega^{10}$ & $\omega^{10}$ & $\omega^{10}$ \\ [0.2em] \hline
$Z_{12}^{\prime}$ & $\omega$ & $\omega^{4}$ & $\omega^{8}$ & $\omega^{5}$ & $\omega^{10}$ & $\omega^{10}$ & $\omega^{3}$ & $\omega^{3}$ & $\omega^{6}$ & $\omega^{7}$ & $\omega^{8}$ & $\omega^{2}$ & $\omega^{11}$ & 1 & $1$ & $\omega^{2}$
\\ \hline   
\end{tabular}
\vspace{-0.in}
\caption{Field content of the $SU(5)\times T^{\prime}$ model. 
The three generations of matter fields in $10$ and 
$\overline{5}$ of $SU(5)$ are in the $T_{3}$, 
$T_{a}$ $(a=1,2)$ and $\overline{F}$ multiplets. 
The Higgses that are needed to generate $SU(5)$ invariant 
Yukawa interactions are $H_{5}$, $H_{\overline{5}}^{\prime}$ 
and $\Delta_{45}$. The flavon fields $\phi$ through 
$\zeta^{\prime}$ are those that give rise to the charged 
fermion mass matrices, while $\xi$, $\eta$, and $S$ 
are the ones that generate neutrino masses.  
The $Z_{12}$ and $Z^\prime_{12}$ charges are given in terms of 
the parameter $\omega = e^{i\pi/6}$.}  
\label{tbl:charge}
\end{table}
%%%%%%%%%%%%%%%%%%%%%%%%%
%
The model includes three right-handed (RH) neutrino fields $N_{lR}$, 
$l=e,\mu,\tau$, which are $SU(5)$ singlets, but are assumed 
to form a triplet of $T^\prime$.
This particle content leads to the 
following Yukawa superpotential up to mass dimension seven:  
%%%%%%%%%%%%%%%%%%%%%%%%%%%%
\begin{equation}
\mathcal{W}_{\mbox{\tiny Yuk}} =  \mathcal{W}_{TT} + \mathcal{W}_{TF} + \mathcal{W}_{\nu} \; ,
\end{equation}
where  
\begin{eqnarray}
\mathcal{W}_{TT} & = & y_{t} H_{5} T_{3} T_{3} + \frac{1}{\Lambda^{2}}  H_{5} \biggl[ y_{ts} T_{3} T_{a} \psi \zeta + y_{c} T_{a} T_{b} \phi^{2} \biggr] + \frac{1}{\Lambda^{3}} y_{u} H_{5} T_{a} T_{b} \phi^{\prime 3} \quad 
\label{eq:Ltt} \; , \\ 
\mathcal{W}_{TF} & = &  \frac{1}{\Lambda^{2}} y_{b} H_{\overline{5}}^{\prime} \overline{F} T_{3} \phi \zeta + \frac{1}{\Lambda^{3}} \biggl[ y_{s} \Delta_{45} \overline{F} T_{a} \phi \psi \zeta^{\prime}  + y_{d} H_{\overline{5}^{\prime}} \overline{F} T_{a} \phi^{2} \psi^{\prime} \biggr]  \quad   
\label{eq:Ltf} \; , \\
\mathcal{W}_{\nu} & = & \lambda_{1} NNS +  \frac{1}{\Lambda^{3}} \biggl[ H_{5}  \overline{F} N \zeta \zeta^{\prime} \biggl( \lambda_{2} \xi + \lambda_{3} \eta\biggr) \biggr] \quad
\label{eq:Lff} \; .
\end{eqnarray}
%%%%%%%%%%%%%%%%%%%%%%%%%%%%%%%
%
The UV completion of these operators is discussed in Ref.~\cite{CM2011-2}. 
Here the parameter $\Lambda$ is the scale 
above which the $T^{\prime}$ symmetry is exact. 
The vacuum expectation values of the flavon 
fields are given by:
%%%%%%%%%%%%%%%%%%%%%%%%%%%%%%%%
\begin{equation}
\left<\xi\right> = \left(\begin{array}{c}
1 \\ 1 \\ 1
\end{array}\right)
\xi_{0} \Lambda \; , \; 
\left< \phi^{\prime} \right> = \left(\begin{array}{c}
1 \\ 1 \\ 1
\end{array}\right) \phi_{0}^{\prime} \Lambda \; , \;  
\left< \phi \right> = \left( \begin{array}{c} 
0 \\ 0 \\ 1
\end{array}\right) \phi_{0} \Lambda \; , \;  
\end{equation}
\begin{equation}
\left< \psi \right> = \left( \begin{array}{c} 1 \\ 0 \end{array}\right)
\psi_{0} \Lambda \; , \; 
\left< \psi^{\prime} \right> = \left(\begin{array}{c} 1 \\ 1 \end{array}\right) \psi_{0}^{\prime} \Lambda \; , 
\end{equation}
\begin{equation}
\left< \zeta \right> = \zeta_{0} \Lambda \; , \; \left< \zeta^{\prime} \right> = \zeta_{0}^{\prime} \Lambda \; , \; 
\left<\eta\right> = \eta_{0} \Lambda \; , \; 
\left<S\right> = s_{0} \Lambda \;  .   
\end{equation}
%%%%%%%%%%%%%%%%%%%%%%%%%%%%%%
%
Note that  all vacuum expectation values are assumed to 
be real and they don't contribute to CP violation. 
The Yukawa couplings are also real.
The reality of the Yukawa 
coupling constants is ensured by the 
presence of  sufficient number of the complex 
flavon fields which allows
to absorb the complex phases in the Yukawa coupling 
constants by phase redefinitions of the fields.

The superpotential gives rise to the following 
mass matrix for the up-type quarks,
%%%%%%%%%%%%%%%%%%%%%%%%%%%%
\begin{equation}
M_{u} = \left( \begin{array}{ccc}
i \phi^{\prime 3}_{0}  & (\frac{1-i}{2}) \phi_{0}^{\prime 3} & 0 \\
(\frac{1-i}{2})  \phi_{0}^{\prime 3}  & \phi_{0}^{\prime 3} + (1 - \frac{i}{2}) \phi_{0}^{2} & y^{\prime} \psi_{0} \zeta_{0} \\
0 & y^{\prime} \psi_{0} \zeta_{0} & 1
\end{array} \right) y_{t}v_{u}, \qquad , 
\end{equation}
%%%%%%%%%%%%%%%%%%%%%%%%%%%
%
and the following down-type quark and charged lepton mass matrices,
%%%%%%%%%%%%%%%%%%%%%%%%
\begin{eqnarray}
\label{Md}
M_{d}  & = & \left( \begin{array}{ccc}
0 & (1+i) \phi_{0} \psi^{\prime}_{0} & 0 \\
-(1-i) \phi_{0} \psi^{\prime}_{0} & \psi_{0} \zeta^{\prime}_{0} & 0 \\
\phi_{0} \psi^{\prime}_{0} & \phi_{0} \psi^{\prime}_{0} & \zeta_{0} 
\end{array}\right) y_{d} v_{d} \phi_{0} \; , \\
\label{Me}
M_{e} & = & \left( \begin{array}{ccc}
0 & -(1-i) \phi_{0} \psi^{\prime}_{0} & \phi_{0} \psi^{\prime}_{0} \\
(1+i) \phi_{0} \psi^{\prime}_{0} & -3 \psi_{0} \zeta^{\prime}_{0} & \phi_{0} \psi^{\prime}_{0} \\
0 & 0 & \zeta_{0} 
\end{array}\right) y_{d} v_{d} \phi_{0} \; . 
\end{eqnarray}
%%%%%%%%%%%%%%%%%%%%%
%
The model describes successfully the quark 
masses and mixing, as discussed in
detail in \cite{Chen:2007afa}.
The presence of the complex elements in $M_{u}$
and $M_{d}$ allows to account for the observed 
CP violation in the quark sector as well \cite{Chen:2009um}.
We concentrate in what follows on the lepton sector.

  In the lepton sector, the superpotential leads to 
the following Dirac neutrino mass matrix,
%%%%%%%%%%%%%%%%%%%%
\begin{equation}
M_{D} = \left( \begin{array}{ccc}
2\xi_{0} + \eta_{0} & -\xi_{0} & -\xi_{0} \\
-\xi_{0} & 2\xi_{0} & -\xi_{0} + \eta_{0} \\
-\xi_{0} & -\xi_{0}+\eta_{0} & 2\xi_{0} 
\end{array}\right) \zeta_{0} \zeta^{\prime}_{0} v_{u} 
% \equiv h_{D} v_{u} \; ,
\equiv \tilde{Y}_{\nu} v_{u} \; ,
\label{MD}
\end{equation}
%%%%%%%%%%%%%%%%%%%%%%%%%%%%
%
and to the following RH neutrino Majorana mass matrix,
%%%%%%%%%%%%%%%%%%%%%%%%%%%%
\begin{equation}
M_{RR} = \left( \begin{array}{ccc}
1 & 0 & 0 \\
0 & 0 & 1 \\
0 & 1 & 0 
\end{array}\right) s_{0} \Lambda \; .
\end{equation}
%%%%%%%%%%%%%%%%%%%%%%%%%%%%%%%%
%

 We note that the complex CG coefficients appear 
in the product rules involving the spinorial 
representations ${\bf 2, \; 2^{\prime}, \; 2^{\prime\prime}}$ 
of $T^\prime$. Because $(T_{1},T_{2})$ transform as 
the spinorial representation ${\bf 2}$, 
the charged fermion mass matrices, 
$M_{u}, \; M_{d}, \; M_{e}$, are complex. 
On the other hand, the neutrinos involve only the 
vectorial-like representations
${\bf 1,  \; 1^{\prime}, 1^{\prime\prime}, 3}$ of  $T^\prime$. 
Therefore the neutrino Dirac and Majorana mass matrices 
are real and thus CP conserving.

Note also that the Dirac neutrino mass matrix, $M_{D}$, 
is real and symmetric. As can be easily shown, 
it is diagonalized by the TBM matrix,
%%%%%%%%%%%%%%%%%%%%%%%%%%%%%%%
\begin{equation}
\label{eq:MDdiag}
% V^{\dagger} K^{1/2} M_{D} K^{1/2} V = 
 U^T_{\mbox{\tiny TBM}}\, M_D\,  U_{\mbox{\tiny TBM}}=
M_{D}^{\mbox{\tiny diag}}\, = 
diag(3\xi_{0} + \eta_{0},\eta_{0},3\xi_{0} - \eta_{0}) 
% \quad \mbox{and} \quad V = U_{\mbox{\tiny TBM}} \; ,
\end{equation} 
%%%%%%%%%%%%%%%%%%%%%%%%%%%%
%
where all elements in the diagonal matrix 
$M_{D}^{\mbox{\tiny diag}}$ are real.
% and $K$ is a diagonal phase matrix. 

The RH neutrino Majorana mass matrix $M_{RR}$ 
is diagonalised by the unitary matrix $S$:
%%%%%%%%%%%%%%%%%%%%%%%%%%%%%%%%%%%
\begin{equation}
S^T\, M_{RR}\, S = D_N = diag(M_1,M_2,M_3) 
= s_0\Lambda\, diag(1,1,1)
\,,~~M_j > 0,~~j=1,2,3\,,
\label{MRRdiag}
\end{equation}
%%%%%%%%%%%%%%%%%%%%%%%%%%%%%%%%%
%
where 
%%%%%%%%%%%%%%%%%%%%%%%%%%%%%%%%%%%%%%%%%
\be S= \left(
  \begin{array}{ccc}
    1 & 0 & 0 \\
    0 & 1/\sqrt{2} & -i/\sqrt{2} \\
   0 & 1/\sqrt{2} & i/\sqrt{2} \\
  \end{array}
\right)\,.
\label{S}
\ee
%%%%%%%%%%%%%%%%%%%%%%%%%%%%%%%%%%%%%%%%
%
and $M_j$ are the masses of the 
heavy Majorana neutrinos $N_j$ (possessing definite masses),
%%%%%%%%%%%%%%%%%%%%%%%%%%%%%
\be
N_j = S^{\dag}_{jl}N_{lR} + S^{T}_{jl}\,C (\bar{N}_{jR})^T = C(\bar{N}_j)^T\,,~
j=1,2,3\,,
\label{Nk}
\ee
%%%%%%%%%%%%%%%%%%%%%%%%%%
%
$C$ being the charge conjugation matrix. 
Thus, to leading order, the masses of the three heavy 
Majorana neutrinos $N_j$ coincide, 
$M_j = s_0\Lambda \equiv M$, $j=1,2,3$.
It follows from eq. (\ref{MRRdiag}) that 
$S^* S^\dagger$ is a real matrix, so 
$S^* S^\dagger = SS^T$.

 The effective Majorana mass matrix 
of the left-handed (LH) 
flavour neutrinos, $M_{\nu}$, 
which is generated by the see-saw mechanism, 
%%%%%%%%%%%%%%%%%%%%
\begin{equation}
M_{\nu} = -M_{D} M_{RR}^{-1} M_{D}^{T} \; ,
\end{equation}
%%%%%%%%%%%%%%%%%%%%%
%
is also diagonalized by the TBM matrix,
%%%%%%%%%%%%%%%%%%%%
\begin{equation}
% \U_{TBM}^{T} M_{\nu} U_{TBM} = 
 U_{TBM}^{T} M_{\nu} U_{TBM} = 
\mbox{diag}( (3\xi_{0} + \eta_{0})^{2}, \eta_{0}^{2}, -(-3\xi_{0}+\eta_{0})^{2}) \frac{(\zeta_{0} \zeta_{0}^{\prime} v_{u})^{2} }{ s_{0}\Lambda} =
\tilde{Q}\,diag(m_1,m_2,m_3)\,\tilde{Q}^T\,.
\label{Mnudiag}
\end{equation}
%%%%%%%%%%%%%%%%%
% 
Here $\tilde{Q}$ is a 
diagonal phase matrix, $\tilde{Q} = diag (1,1,\pm i)$,
and $m_k > 0$, $k=1,2,3$, are the masses of the three 
light Majorana neutrinos:
%%%%%%%%%%%%%%%%%%%%%%%%%%%%%%%%%%%%
\be
m_1 \equiv \frac{(X+3Z)^2}{M},
\quad
m_2 \equiv \frac{X^2}{M},
\quad m_3 \equiv
\frac{(X-3Z)^2}{M}\,,
\label{masses}
\ee
%%%%%%%%%%%%%%%%%%%%%%%%%%%%%%%%%%
%
where $X \equiv \eta_0(\zeta_{0} \zeta_{0}^{\prime} v_{u})$
and $Z \equiv \xi_0 (\zeta_{0} \zeta_{0}^{\prime} v_{u})$.
It follows from eqs. (\ref{UeUnu1}) and 
(\ref{Mnudiag}) that 
$\tilde{U}_\nu\,Q =  U_{TBM}\,\tilde{Q}$.
Thus, we have $Q = \tilde{Q}$ 
and the Majorana phases $\alpha_{21}$ and 
$\alpha_{31}$ are predicted (to leading order) 
to have CP conserving values:
$\alpha_{21} = 0$ and $\alpha_{31} = \pm \pi$.

  One special property of $M_{\nu}$ is that it is form 
diagonalizable~\cite{Chen:2009um}. In other words, 
regardless of the values of $\xi_{0}$ and $\eta_{0}$, 
$M_{\nu}$ is always diagonalized by the 
TBM matrix, $U_{\mbox{\tiny TBM}}$. 

  The charged lepton mass matrix $M_e$, eq. (\ref{Me}), 
is diagonalised, in general, by the by-unitary transformation:
$M_e = U_{e}M_e^{d}V^\dagger_{eR}$, where $V_{eR}$ and $U_e$ are unitary 
matrices and $M_e^{d} = diag(m_e,m_{\mu},m_{\tau})$, 
$m_l$ being the mass of the charged lepton $l$, $l=e,\mu,\tau$.
Thus, the matrix $U_e$, which enters into the 
expression for the PMNS matrix,
%  $U_{PMNS} = U^\dagger_eU_{\nu}$, 
$U = U^\dagger_eU_{\nu}$,
digonalises the matrix $M_eM^\dagger_e$: 
$U^\dagger_e (M_e\,M^\dagger_e)U_e = (M_e^{d})^2$.
The $3\times 3$ unitary matrix $U_e$ of interest can be 
parametrised, in general, as 
$U_e = \Phi\, V_e(\theta^e_{12},\theta^e_{13},\theta^e_{23},\delta^e)P$,
where   $V_e(\theta^e_{12},\theta^e_{13},\theta^e_{23},\delta^e)$
has the same form as the matrix $V$ in eq. (\ref{V}) 
(with $\theta_{ij}$ and $\delta$  replaced by $\theta^e_{ij}$ 
and $\delta^e$) and $\Phi$ and $P$ are diagonal phase matrices 
containing two and three phases, respectively. 
The phases in $P$ are unphysical and we are not 
going to consider them further. 

   Due to the $SU(5)$ symmetry, the matrices $M_d$ and $M_e$ 
depend on the same three real parameters and two phases, the latter 
having the values $\pm \pi/4$. The three real parameters can be 
fixed by requiring that $M_d$ and $M_e$ reproduce correctly 
the down-quark and charged lepton mass ratios 
$m_d/m_s$, $m_e/m_\mu$, etc., the Cabibbo angle $\theta_c$ and 
other quark mixing observables. The model accounts successfully for 
the quark masses and mixing and for the charged lepton 
masses ~\cite{Chen:2007afa,Chen:2009gf}.
In particular, the well-known relations 
$\sin\theta_c \cong \sqrt{m_d/m_s}$, $m_e/m_{\mu}\cong m_d/(9m_s)$, 
etc. are fulfilled.

 Fitting the indicated quark sector observables and charged 
lepton masses one finds that \cite{Chen:2009gf} 
two of the three angles in the matrix 
$U_e$ are extremely small, 
%%%%%%%%%%%%%%%%%%%%%%%%%%%%
\be 
\sin\theta^e_{13} \cong 1.3\times 10^{-5}\,,~ 
\sin\theta^e_{23} \cong 1.5\times 10^{-4}\,,
\label{the12the23}
\ee
%%%%%%%%%%%%%%%%%%%%%%%%%%%%%%
%
while the third satisfies:
%%%%%%%%%%%%%%%%%%%
\be 
\sin\theta^e_{12} = \frac{1}{3}\,\sin\theta_{c}\,.
\label{the12}
\ee
%%%%%%%%%%%%%%%%%%%%
%
It follows from the quoted results that, to a very good 
approximation, we can set $\theta^e_{13}= \theta^e_{23} = 0$ 
in the expression for $U_e$. In this approximation we have 
$U_e = \Phi\,R_{12}(\theta^e_{12})$, where 
$\Phi = diag(1,e^{i\varphi},1)$ and
%%%%%%%%%%%%%%%%%%%%%%%%%%%%%%%%%%%%%%%%%
\be  
R_{12}(\theta^e_{12}) = 
\left(
  \begin{array}{ccc}
    \cos\theta^e_{12} & \sin\theta^e_{12} & 0 \\
    -\,\sin\theta^e_{12} &  \cos\theta^e_{12}  & 0 \\
   0 & 0 & 1 \\
  \end{array}
\right)\,.
\ee
%%%%%%%%%%%%%%%%%%%%%%%%%%%%%%%%%%%%%%%%
%
Comparing the expressions in the left-hand and right-hand 
sides of the equation $M_e\,M^\dagger_e = U_e(M_e^{d})^2U_e^\dagger$
and assuming, without loss of generality, that 
$\phi_{0} \psi^{\prime}_{0}\psi_{0} \zeta^{\prime}_{0} > 0$ and 
$\cos\theta^e_{12}\sin\theta^e_{12} >0$, we find that
%%%%%%%%%%%%%%%%%%%%%%%%
\be
\varphi = \frac{\pi}{4}\,.
\label{varphi}
\ee
%%%%%%%%%%%%%%%%%%%%%
%
In the approximation we are using the PMNS matrix is given  by:
%%%%%%%%%%%%%%%%%%%%%%%%%%%%%%%%%%%%%
\begin{eqnarray}
\label{PMNSTBM1}
% U^{TBM}_{PMNS} 
U^{SU(5)\times T^\prime} &\equiv& U^{\prime}
\cong R^T_{12}(\theta^e_{12})\, \Phi^*(\varphi) \,
U_{TBM}\,Q\,\\
&=& \left(\begin{array}{ccc}
\sqrt{2/3}c^e_{12} + \sqrt{1/6}s^e_{12}e^{-i\varphi} & 
\sqrt{1/3}(c^e_{12} - s^e_{12}e^{-i\varphi}) & 
             \sqrt{1/2}s^e_{12}e^{-i\varphi} \\
  \sqrt{2/3}s^e_{12} - \sqrt{1/6}c^e_{12}e^{-i\varphi} & 
\sqrt{1/3}(s^e_{12} + c^e_{12}e^{-i\varphi}) &
 -\sqrt{1/2} c^e_{12}e^{-i\varphi}\\
-\sqrt{1/6} & \sqrt{1/3} & \sqrt{1/2}
\end{array}\right)\,Q \;,
 \label{PMNSTBM2}
\end{eqnarray}
%%%%%%%%%%%%%%%%%%%%%%%%%%%%%%%%%%%%
%
where $c^e_{12} = \cos\theta^e_{12}$, 
$s^e_{12} = \sin\theta^e_{12}$.
It follows from the above expression and 
eqs. (\ref{UPMNS}) - (\ref{Q}) that
% ~\cite{FPR,HPR07,Chen:2009gf,Marzocca:2011dh}
~\cite{FPR,HPR07,Marzocca:2011dh}
up to corrections of the order of $\sin^2\theta^e_{12}$,
$\theta_{23}$ takes its TBM value, $\theta_{23} = \pi/4$,
and that to leading order in $\sin\theta^e_{12}$ we have
%%%%%%%%%%%%%%%%%%%%%%%%%
\be 
\sin\theta_{13} \cong  \frac{1}{\sqrt{2}}\,\sin\theta^e_{12}\,, 
% =  \frac{1}{3\sqrt{2}}\,\sin\theta_{c}\,,
\label{th13the12}
\ee
%%%%%%%%%%%%%%%%%%%%%%%%%
%
and
%%%%%%%%%%%%%%%%%%%%%%%%%
\be 
\sin^2\theta_{12} \cong \frac{1}{3} -
 \frac{2}{3}\,\sin\theta^e_{12}\,\cos\varphi =
\frac{1}{3} - \frac{2\sqrt{2}}{3}\,\sin\theta_{13}\,\cos\varphi\,. 
% = \frac{1}{3} - \frac{\sqrt{2}}{9}\,\sin\theta_{c}\,, 
\label{th12th13}
\ee
%%%%%%%%%%%%%%%%%%%%%%%%%
%
% where we have used eqs. (\ref{the12}) and (\ref{varphi}).
Using  eqs. (\ref{the12}) and (\ref{varphi}) we get 
~\cite{Chen:2009gf}:
%%%%%%%%%%%%%%%%%%%%%%%%%%%%%%%%%%%%%%%%%%%%%%%%
\be 
\sin\theta_{13} \cong \frac{1}{3\sqrt{2}}\,\sin\theta_{c}\,,
\label{th13}
\ee
%%%%%%%%%%%%%%%%%%%%%%%%%
%
and
%%%%%%%%%%%%%%%%%%%%%%%%%
\be 
\sin^2\theta_{12} \cong 
\frac{1}{3} - \frac{\sqrt{2}}{9}\,\sin\theta_{c}\,. 
\label{th12}
\ee
%%%%%%%%%%%%%%%%%%%%%%%%%
%
Thus, in the model considered 
the CHOOZ angle $\theta_{13}$ is predicted to be
rather small: using $\sin\theta_c = 0.22$ we get 
from eq. (\ref{th13}), $\sin\theta_{13} = 0.052$.
From a numerical analysis in which the higher order 
corrections were also included one finds \cite{Chen:2009gf}  
$\sin\theta_{13} \cong 0.058$.
This value lies in the 3$\sigma$ interval of 
allowed values of $\sin\theta_{13}$, 
determined in the global analysis  \cite{Fogli:2011qn}
of the neutrino oscillation data.
The correction to the TBM value of $\sin^2\theta_{12}$ 
given in eq. (\ref{th12}), is negative.
The value of $\sin^2\theta_{12} \cong 0.299$ predicted by the model, 
lies within the $1\sigma$ allowed range,
found in the global data analysis \cite{Fogli:2011qn}.

  It is possible to relate also 
the Dirac CP violating phase $\delta$, present in 
% $U_{PMNS}$, 
$U$, eqs. (\ref{UPMNS}) - (\ref{Q}), 
with the phase $\varphi$ in 
% $U^{TBM}_{PMNS}$. 
$U^{\prime}$.
It proves convenient first to multiply  
the elements of the first row and of the first and 
of the second columns of the PMNS matrix 
in eq. (\ref{PMNSTBM2}) by (-1), and the elements 
of the second row by $(-e^{i\varphi})$.
Multiplying by (-1) ((-$e^{i\varphi}$))
the first (second) row is equivalent of redefining 
the phase of the electron (muon) field in 
the weak charged current.
Changing the signs of the elements 
of the first and of the second columns of 
the PMNS matrix in eq. (\ref{PMNSTBM2})
can be compensated by multiplying the matrix $Q$ 
containing the two Majorana phases by 
$diag(-1,-1,1)$: 
$diag(-1,-1,1)Q = 
diag(-1,-1,\pm i) = (-1)diag(1,1,\mp i) = 
(-1)Q^*$, 
where the overall factor (-1) in the 
matrix $Q^*$ has no 
physical significance and will be dropped 
in our further discussions.
After this simple manipulations 
the matrix in eq. (\ref{PMNSTBM2}) 
takes a form which is similar to that 
of the standard parametrisation of the PMNS matrix, 
eqs. (\ref{UPMNS}) - (\ref{Q}):
%%%%%%%%%%%%%%%%%%%%%%%%%%%%%%%%%%%%%
\begin{eqnarray}
\label{PMNSTBM1}
% U^{TBM}_{PMNS} 
% U^{SU(5)\times T^\prime} &\equiv& 
U^{\prime} &\cong& \left(\begin{array}{ccc}
\sqrt{2/3}c^e_{12} - \sqrt{1/6}s^e_{12}e^{-i\varphi'} & 
\sqrt{1/3}(c^e_{12} + s^e_{12}e^{-i\varphi'}) & 
             \sqrt{1/2}s^e_{12}e^{-i\varphi'} \\
 - \sqrt{1/6}c^e_{12} -\sqrt{2/3}s^e_{12}e^{i\varphi'}  & 
\sqrt{1/3}(c^e_{12} - s^e_{12}e^{i\varphi'}) &
 \sqrt{1/2} c^e_{12}\\
\sqrt{1/6} & - \sqrt{1/3} & \sqrt{1/2}
\end{array}\right)\,Q^* \;,
 \label{PMNSTBM3}
\end{eqnarray}
%%%%%%%%%%%%%%%%%%%%%%%%%%%%%%%%%%%%
%
where 
%%%%%%%%%%%%%%%%%%%%%%%%
\be
\varphi' = \frac{\pi}{4} \pm \pi\,.
\label{varphip}
\ee
%%%%%%%%%%%%%%%%%%%%%
%
We note that the phases of the 
% $(U^{TBM}_{PMNS})_{e1}$ and  $(U^{TBM}_{PMNS})_{e2}$ 
% elements of $U^{TBM}_{PMNS}$, 
$U^{\prime}_{e1}$ and  $U^{\prime}_{e2}$ 
elements of $U^{\prime}$, 
$\kappa_{e1}$ and $\kappa_{e2}$, are exceedingly small:
$\kappa_{e1} \cong s^e_{12}\sin\varphi'/2 
\cong (- 0.0259) \cong (-1.5^\circ)$, 
$\kappa_{e2} \cong - s^e_{12}\sin\varphi'
\cong 0.0519 \cong (3.0^\circ)$.
% Thus, the imaginary parts of  $(U^{TBM}_{PMNS})_{e1}$ and  
% $(U^{TBM}_{PMNS})_{e2}$ are much smaller than  their real parts 
% and we have  $(U^{TBM}_{PMNS})_{ek} \cong |(U^{TBM}_{PMNS})_{ek}|$, 
% $k=1,2$. 
 Thus, the imaginary parts of  $U^{\prime}_{e1}$ and  
$U^{\prime}_{e2}$ are much smaller than  their real parts 
and we have  $U^{\prime}_{ek} \cong |U^{\prime}_{ek}|$, 
$k=1,2$.

  Comparing the real and imaginary parts of the 
quantity $U^*_{e1}U_{\mu 1}U_{e3}U^*_{\mu3}$, 
calculated using  eqs. (\ref{UPMNS}) - (\ref{Q}),
with those obtained utilizing eq. (\ref{PMNSTBM3}) 
(see, e.g., \cite{Marzocca:2011dh}) and assuming that 
the Dirac phase $\delta$ lies in the ``standard''
interval $[0,2\pi]$ we find
%%%%%%%%%%%%%%%%%%%%%%%%
\be
\delta = \varphi' = \varphi + \pi \cong 
% \frac{\pi}{4} + \pi = 
\frac{5}{4}\,\pi.
\label{delta}
\ee
%%%%%%%%%%%%%%%%%%%%%
%
Note that the sign of $\cos\delta$ and the value 
of $\delta$ are compatible with those suggested 
by the current neutrino oscillation data.
Substituting $\varphi$ with $(\delta - \pi)$ 
in eq. (\ref{th12}) and using 
the results obtained in \cite{Marzocca:2011dh} on 
the values of $\cos\delta$, 
allowed by the existing data on 
$\sin^2\theta_{12}$ and $\sin\theta_{13}$,  
we get for $\sin\theta_{13} = 0.058$, 
predicted by the model, and the $3\sigma$ ($2\sigma$) 
interval of experimentally allowed values of 
$\sin^2\theta_{12}$:
%%%%%%%%%%%%%%%%%%%%%%%%
\be
-1 \leq \cos\delta \ltap 0.4~(0.1)\,.
\label{cosdelta}
\ee
%%%%%%%%%%%%%%%%%%%%%
%
Thus, in the model considered, $\delta = 0$ and, more generally,
the values of  $\cos\delta$ from the interval 
$0.4 <  \cos\delta \leq 1$, are excluded at $3\sigma$.
 
  As is well known,
the quantity $J_{CP} = {\rm Im}(U^*_{e1}U_{\mu 1}U_{e3}U^*_{\mu3})$
is the rephasing invariant associated  with 
the Dirac CP violating phase $\delta$ in the PMNS matrix.
It determines the magnitude of CP violation effects in 
neutrino oscillations \cite{PKSP3nu88} and
is analogous to the rephasing invariant associated with the 
Dirac phase in the Cabibbo-Kobayashi-Maskawa
quark mixing matrix, introduced in \cite{CJ85}. 
In the model considered the rephasing invariant $J_{CP}$ 
is given to leading order in  $\sin\theta^e_{12}$ by
%%%%%%%%%%%%%%%%%%%%
\be
J_{CP} \cong \frac{1}{6}\,\sin\theta^e_{12}\, \sin\delta 
\cong \frac{1}{18}\,\sin\theta_{c}\, \sin\delta 
= -\, \frac{1}{18\sqrt{2}}\,\sin\theta_{c}\,.  
% \cong - 8.6\times 10^{-3}\,.
\label{JCPTp}
\ee
%%%%%%%%%%%%%%%%%%%%%%%%%%%%%

Finally, we give the expression for the 
PMNS matrix 
% $U^{TBM}_{PMNS}$ 
$U^{\prime}$
obtained numerically in \cite{Chen:2009gf}, 
in which the higher order correction in  
$\sin\theta^e_{12}$, as well as   
the corrections due to the nonzero values of 
$\sin\theta^e_{13}\sim 10^{-5}$ and   $\sin\theta^e_{23}\sim 10^{-4}$,
are all taken into account: 
%%%%%%%%%%%%%%%%%%%%%%%%%%%%%%%%%%%%
%\be
% U^{TBM}_{PMNS}\simeq\left(
% \begin{array}{ccc}
%  0.838e^{-i178.6^{\circ}}  & 0.543e^{-i173.5^{\circ}} &
% 0.0582e^{i138.2^{\circ}} \\
% 0.363e^{-i3.53^{\circ}}  & 0.610e^{-i 172.8^{\circ}} &  
% 0.705e^{i 4^{\circ}} \\
% 0.408 e^{i180^{\circ}}  & 0.577 & 0.707
% \end{array}
% \right).
% \label{PMNSnumeric}
% \ee
%%%%%%%%%%%%%%%%%%%%%%%%%%%%%%%%%%%%
\be
% U^{TBM}_{PMNS}
U^{SU(5)\times T^\prime} \equiv U^{\prime}
\simeq\left(
\begin{array}{ccc}
 0.838e^{i1.4^{\circ}}  & 0.543e^{i6.5^{\circ}} & 0.058e^{-i221.8^{\circ}} \\
-\,0.363e^{-i3.53^{\circ}}  & 0.610e^{i7.2^{\circ}} &  0.705e^{i 4^{\circ}} \\
0.408   & -\,0.577 & 0.707
\end{array}
\right)\,Q.
\label{PMNSnumeric}
\ee
%%%%%%%%%%%%%%%%%%%%%%%%%%%%%
%
It follows from this expression of 
% $U^{TBM}_{PMNS}$
$U^{\prime}$
that the precise values of 
$\sin\theta_{13}$, $\delta$ and $J_{CP}$,
predicted by the model, read: 
$\sin\theta_{13} = 0.058$, 
$\delta = 221.8{\circ}$ and  
$J_{CP} = -9.66\times 10^{-3}$.
The predictions of the $SU(5)\times T^\prime$ 
model of interest for $\sin\theta_{13}$, 
$\sin^2\theta_{12}$, $\delta$ and $J_{CP}$ 
can be tested directly in the upcoming 
neutrino oscillation experiments.

% The charged
% lepton mixing matrix $V_{e,L}$
% used in our analysis is given by:
% %%%%%%%%%%%%%%%%%%%%%%%%%%%%%%%%%%%%%%%
% \be
% U_{e}= \left(\begin{array}{ccc}
%  0.997 e^{i 177^{\circ}} &0.0823 e^{i 131^{\circ}}  & 1.31\times 10^{-5} e^{-i 45^{\circ}} \\
% 0.0823 e^{i 41.8^{\circ}}  & 0.997 e^{i 176^{\circ}} & 0.000149 e^{-i 3.58^{\circ}} \\
%  1.14\times 10^{-6} & 0.000149 & 1
% \end{array}\right)
% \ee
% %%%%%%%%%%%%%%%%%%%%%%%%%%%%%%%
%

%%%%%%%%%%%%%%%%%%%%%%%%%%%%
%
\section{Neutrino Masses, Spectrum and the $\betabeta$-Decay 
Effective Majorana Mass}
%
%%%%%%%%%%%%%%%%%%%%%%%%%%%

 In the model considered we have $\cos2\theta_{12} > 0$ and 
therefore $\Delta m^2_{21}$ should also be positive.
It is easy to show then that 
% in the model considered
we also have $\Delta m^{2}_{A} \equiv \Delta m^2_{31} > 0$, 
i.e., that in this model only light neutrino mass spectrum with 
{\it normal ordering} (or \emph{normal hierarchy}) is possible.
Indeed, using the expressions given in (\ref{masses})
one finds that there do not exist values of $X$ and $Z$
for which  the condition which defines the spectrum with
{\it inverted ordering}, namely,
$\Delta m^{2}_{A} \equiv \Delta m^2_{31} < 0$,
or  $m_2 > m_1 > m_3$, is satisfied.

  The neutrino masses in the model considered, 
as it follows from eq. (\ref{masses}), 
depend on two parameters: 
$X/\sqrt{M}$ and $Z/\sqrt{M}$. The latter can be 
determined, e.g., using the data on 
% the values of 
$\Delta m^{2}_{21}$ and $\Delta m^2_{31}$.
Thus, the absolute values of the three light neutrino 
masses $m_{1,2,3}$ in the $SU(5)\times T^\prime$ model 
under investigation are fixed 
by the values of $\Delta m^{2}_{21}$ and $\Delta m^2_{31}$.

  It proves convenient to use $\Delta m^2_{21}$
and the ratio $r \equiv \Delta m^{2}_{21}/\Delta m^2_{31}$,
instead of  $\Delta m^{2}_{31}$,
to determine the masses $m_{1,2,3}$.
We have for the $3\sigma$ allowed range
of the indicated ratio:
%%%%%%%%%%%%%%%%%%%%%%%%%%%%%%%%%
\be
r \equiv \frac{\Delta m^{2}_{\odot}}{\Delta m^{2}_{A}} 
\equiv  \frac{\Delta m^{2}_{21}}{\Delta m^2_{31}} 
= \frac{X^4-(X+3Z)^4}{(X-3Z)^4-(X+3Z)^4}= 0.032 \pm
0.006,
\label{r}
\ee
%%%%%%%%%%%%%%%%%%%%%%%%%%%%%%%%%%%%
%
Using the fact that, e.g., $m^2_2 = 
\Delta m^{2}_{21} + m^2_1$ 
and eq. (\ref{masses}), we can express 
the heavy Majorana neutrino mass $M$ 
in terms of  $\Delta m^{2}_{21}$, $X$ and $Z$:
$M^2 = (X^4 - (X + 3Z)^4)/(\Delta m^{2}_{21})$.
Substituting this  expression for 
$M$ in eq. (\ref{masses}), we get for $m_i^2$:
%%%%%%%%%%%%%%%%%%%%%%%%%%%%%%%%%%%%%%%%%%%%%
\be m_1^2= \frac{\Delta m^{2}_{21}}{\left(\dfrac{X}{X+3Z}
\right)^4-1},
\quad m_2^2= \frac{\Delta m^{2}_{21}}{1-\left(\dfrac{X+3Z}{X} \right)^4},
\quad m_3^2= \frac{\Delta m^{2}_{21}}{\left(\dfrac{X}{X-3Z}\right)^4-\left(\dfrac{X+3Z}{X-3Z}
\right)^4}.
\label{recast}
\ee
%%%%%%%%%%%%%%%%%%%%%%%%%%%%%%%%%%
%
Given $r$, eq. (\ref{r}) implies
a relation between the parameters $X$ and $Z$.
As can be shown, to first order in $r$, 
this relation reads: $3Z\simeq (20 r -2)X$.
A numerical analysis performed by
us showed that the values of the light neutrino
masses, calculated using 
eqs. (\ref{masses}) and (\ref{r}) 
and the values of  
$\Delta m^{2}_{21}$ and $r$ as input,
are reproduced with a remarkable precision
when calculated employing the approximate 
relation between  $Z$ and $X$ given above 
instead of the exact one implied by eq. (\ref{r}).
Utilizing the relation  $3Z\simeq (20 r -2)X$ and eq. 
(\ref{recast}) allows to express $m^2_{1,2,3}$ 
as simple functions of $\Delta m^{2}_{21}$ and $r$:
%%%%%%%%%%%%%%%%%%%%%%%%%%%%%%%%%%%%%%%%%%
\be
m_1^2 \cong \frac{\Delta m^{2}_{21}(20r-1)^4}{1 - \left( 20r - 1\right)^4},
\quad m_2^2\cong \frac{\Delta m^{2}_{21}}{1-\left(20 r-1
\right)^4},
\quad m_3^2\cong \frac{\Delta m^{2}_{21} (20r-3)^4}
{1 - (20r-1)^4}.
\label{massr}
\ee
%%%%%%%%%%%%%%%%%%%%%%%%%%%%%%%%%%%%%%%%%
%
We also have:
%%%%%%%%%%%%%%%%%%%%%%%
\be
M\,\sqrt{\Delta m^{2}_{21}} \cong X^2
\left (1 - (20r -1)^4\right)^{\frac{1}{2}} \cong X^2\,,
\label{MX}
\ee
%%%%%%%%%%%%%%%%%%
%
where we have neglected $(20r -1)^4\cong 0.017$. 
For, e.g., $M=10^{12}$ GeV, using 
$X = \eta_0(\zeta_{0} \zeta_{0}^{\prime} v_{u})$, 
$v_u = 174$ GeV and
$\Delta m^{2}_{21}= 7.59\times 10^{-5}~{\rm eV^2}$ 
we obtain:  
$(\eta_0\zeta_{0} \zeta_{0}^{\prime})^2 \cong 2.89\times 10^{-4}$.
Taking into account that  
$Z \equiv \xi_0 (\zeta_{0} \zeta_{0}^{\prime} v_{u})$, we find 
also $\xi_0 \cong (20r - 2)\eta_0/3 \cong - 0.453\eta_0$.
For  $M=10^{12}$ GeV we have: 
$\xi_0 \zeta_{0} \zeta_{0}^{\prime}\cong -7.74\times 10^{-3}$.

In Fig. \ref{fig:spectrum} the light neutrino 
masses $m_{1,2,3} > 0$ are plotted as functions 
of the ratio $r$, with $\Delta m^{2}_{21}$ set to its 
best fit value.
As Fig. \ref{fig:spectrum} indicates, 
in the model with approximate $SU(5)\times T^\prime$ symmetry
under study, the light neutrino masses are 
allowed to vary (due to the uncertainties in the 
experimentally determined values of 
$\Delta m^{2}_{21}$ and $r$) 
in rather narrow intervals around the values
$m_1=1.14 \times 10^{-3}eV$,  $m_2=8.78 \times
10^{-3}eV$,  $m_3=4.89 \times 10^{-2}eV$.
This, implies, in particular that the model
provides specific predictions for the sum of the
light neutrino masses as well as well for the effective
Majorana mass in neutrinoless double beta
($\betabeta$-decay), $\meff$.

For the sum of the three light neutrino masses we obtain:
%%%%%%%%%%%%%%%%%%%%%%%%%%%%%%%%%%%%%%
\be 
m_1 + m_2 + m_3\cong \left(\frac{\Delta m^{2}_{21}}
{1-(20r-1)^4}\right )^{\frac{1}{2}}
\left [ 1 + (20r-1)^2 + (20r-3)^2\right ]   
\label{summia}
\ee
%%%%%%%%%%%%%%%%%%%%%%%%%%%%%%%%%%%%%%%%
%
Numerically we get using the best fit values of 
$\Delta m^{2}_{21}$ and $r$:
%%%%%%%%%%%%%%%%%%%%%%%%%%%%%%%%%%%%
\be
m_1 + m_2 + m_3 \cong 5.9\times 10^{-2}~{\rm eV}\,,
\ee
%%%%%%%%%%%%%%%%%%%%%%%%%%%%%%%%%%%
%

To leading order in $\sin\theta^e_{12}$, 
the $\betabeta$-decay effective Majorana mass 
(see, e.g., \cite{BiPet87}),
$\meff \equiv |\sum_{k=1}^3 (U_{PMNS})^2_{ek}\,m_k|$, 
is given by:
%%%%%%%%%%%%%%%%%%%%%%%%%%%%%%%
\begin{eqnarray}
\meff \cong
 \left | \frac{1}{3}\,(2m_1 + m_2) - 
\frac{2}{3}\,(m_2 - m_1)\,\sin\theta^e_{12}e^{-i\varphi} + 
\frac{1}{2}\,m_3\,(\sin\theta^e_{12})^2\,e^{-i(2\varphi - \pi)}\right |\,, 
\label{meff}
\end{eqnarray}
%%%%%%%%%%%%%%%%%%%%%%%%%%%%%
%
where we have employed the expression for $U_{PMNS}$ 
given in eq. (\ref{PMNSTBM2}).
Using the numerical values of $m_{1,2,3}$ corresponding 
to $r=0.032$ quoted above, eqs. (\ref{the12}) and (\ref{varphi}) and the 
value of $\sin\theta_c = 0.22$, we find that the term
$\propto i\sin\varphi$ and that $\propto m_3$ in  
$\meff$ give  negligible contributions and that
%%%%%%%%%%%%%%%%%%%%%%%%%%%%%
\be
\meff \cong 3.4\times 10^{-3}~{\rm eV}\,.
\ee
%%%%%%%%%%%%%%%%%%%%%%%%%%%%%%%%

%%%%%%%%%%%%%%%%%%%%%%%%%%%%%%%%%%%%%%%%
\begin{figure}[t!]
   \begin{center}
 \includegraphics[width=8cm]{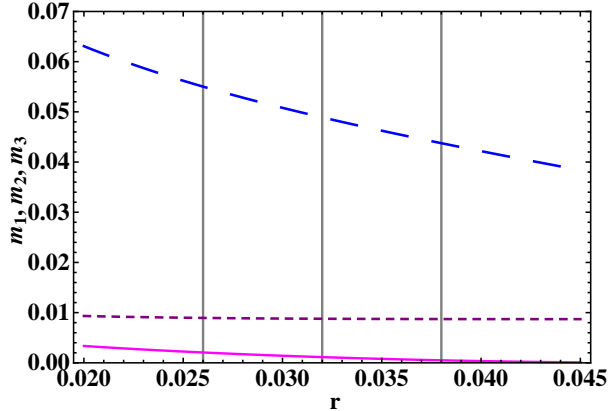}
     \end{center}
   \caption{\label{fig:spectrum} The light neutrino masses
 $m_{1}$ (solid line), $m_2$ (dashed line) and $m_3$ 
(long-dashed line) as functions of the parameter $r$ 
for $\Delta m^2_{21} =7.58\times 10^{-5}~{\rm eV^2}$. 
The three vertical lines correspond,
from left to right, to  $r= 0.026, 0.032, 0.038$, i.e., 
to the $3\sigma$ minimal, best fit and  $3\sigma$ maximal
values of $r$. For $r= 0.032$ we have: 
$m_1=1.14 \times 10^{-3}$ eV,  $m_2=8.788 \times
10^{-3}$ eV,  $m_3=4.89 \times 10^{-2}$ eV. }
\label{fig:spectrum}
\end{figure}
%%%%%%%%%%%%%%%%%%%%%%%%%%%%%%
%

%%%%%%%%%%%%%%%%%%%%%%%%%%%%
%
\section{The LFV Decays $\ell_{i} \rightarrow\ell_j +\gamma$} 

\subsection{The Decays $\ell_{i} \rightarrow\ell_j +\gamma$ 
in SUSY theories}
%
%%%%%%%%%%%%%%%%%%%%%%%%%%%

In the minimal extended Standard Theory with heavy Majorana
right-handed neutrinos, the simultaneous presence of neutrino and
lepton Yukawa couplings, $Y_\nu$ and $Y_e$, leads to lepton flavour
violation. In this scenario, the LFV decay rates and cross sections 
are strongly suppressed by the ratio 
\cite{Petcov:1976ff,Bilenky:1977du}
$|(\sum_{j=2,3}U^*_{\mu j}U_{ej} \Delta m^2_{j1})/M^2_W|^2$,
$M_W\cong 80$ GeV being the $W^\pm$-boson mass, 
leading to, e.g., BR($\mu \rightarrow e \gamma$)$< 10^{-47}$.
This renders the LFV decays and reactions unobservable 
in the ongoing and future planned experiments. 
The situation is the same in the non-SUSY type I see-saw
models in which the heavy Majorana neutrino masses 
$M_k$ are only by few to several orders of magnitude smaller 
than the GUT scale  $M_X\approx 2\times 10^{16}$ GeV 
\cite{Cheng:1980tp,footn4}.

The present experimental upper bounds on the rates 
of the LFV decays $\ell_{i} \rightarrow\ell_j +\gamma$, 
$m_{\ell_i} > m_{\ell_j}$,  $\ell_{1}\equiv e$,  
$\ell_{2}\equiv \mu$, $\ell_{3}\equiv \tau$,
are given by 
\cite{MEG11,Nakamura:2010zzi}
%%%%%%%%%%%%%%%%%%%%%%%%%%%%%%%%%%%%%%%%%%
\be 
% BR(\mu \rightarrow e \gamma)< 1.2 \times 10^{-11},
BR(\mu \rightarrow e \gamma)< 2.4 \times 10^{-12},
\quad BR(\mu
\rightarrow 3 e )< 1 \times 10^{-12},
\quad BR(\tau \rightarrow
\mu \gamma)< 4.4\times 10^{-8}. 
\ee
%%%%%%%%%%%%%%%%%%%%%%%%%%%%%%%%%%%%%%%%%%
%
The first was obtained recently in the MEG experiment 
at PSI and is an improvement by a factor of 5 of the 
previous best upper limit of the MEGA experiment, 
published in 1999 \cite{Brooks:1999pu}.
The projected sensitivity of the 
MEG experiment is \cite{MEG11}:
%%%%%%%%%%%%%%%%%%%%%%%
\be 
BR(\mu \rightarrow e \gamma)\sim  10^{-13}.
\ee 
%%%%%%%%%%%%%%%%%%%%%%%%%%%%%%%%%%%%%%%%
\begin{figure}[t!]
   \begin{center}
\includegraphics[width=7.5cm]{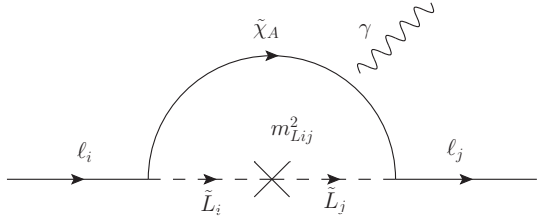}
     \end{center}
   \caption{ Feynman diagrams of the dominant contribution to the 
$\ell_i \rightarrow\ell_j +\gamma $ radiative decay amplitude 
in the mass insertion approximation. $\tilde\chi_A$ corresponds to charginos 
or neutralinos and $\tilde L_i$ are the slepton doublets. 
The photon can be emitted from the chargino or the slepton lines. 
\label{fig:Feynman}}
\end{figure}
%%%%%%%%%%%%%%%%%%%%%%%%%%%%%%
%

 In the supersymmetric (SUSY) theories these decay rates 
can be largely enhanced due to contributions from the slepton
part of the soft SUSY breaking Lagrangian, $\mathcal{L}_{soft}$:
%%%%%%%%%%%%%%%%%%%%%%%
\ba 
-\mathcal{L}_{soft}&=& (m^2_{\tilde L})_{ij}\tilde{L}^\dagger_i
\tilde{L}_j +(m^2_{\tilde e})_{ij} \tilde{e}^\ast_{Ri}\tilde{e}_{Rj}
+ (m^2_{\tilde \nu})_{ij} \tilde{\nu}^\ast_{Ri}\tilde{\nu}_{Rj}\nn\\
&& + \left( (A_e)_{ij}H_d \tilde{e}^\ast_{Ri}\tilde{L}_j+
(A_\nu)_{ij}H_u\tilde{\nu}^\ast_{Ri}\tilde{L}_j+ h.c. \right),
\ea
%%%%%%%%%%%%%%%%%%%%%%%
%
where  $m^2_{\tilde L}$ and $m^2_{\tilde e}$ are the left-handed (LH) and
right-handed (RH) charged slepton mass matrices, respectively,
$m^2_{\tilde \nu}$ is the right-handed sneutrino soft mass term,
$A_e$ and $A_\nu$ are trilinear couplings and
$H_d$ and $H_u$ are the two Higgs doublet 
fields present in the SUSY theories. 
% and $\tilde{L}_j$, $\tilde{e}_{Rj}$ and $\tilde{\nu}_{Rj}$, $j=e,\mu,\tau$,
% are the LH, RH charged slepton and RH sneutrino fields, respectively.
Non-zero off-diagonal elements in the slepton mass matrix 
could induce lepton flavour violation, but only relatively 
small values could satisfy the upper
bounds quoted above. Soft-breaking terms are, however, 
subjected to renormalization through Yukawa and gauge interactions 
in such a way that LFV is induced in the slepton mass matrix at 
``low energies''. Indeed, in addition to a LF conserving part, 
the renormalisation group (RG) equation for the left 
handed slepton mass matrix present, in general, 
off-diagonal terms which are a  source of LFV. 
 
  The indicated generic possibility is realized in the SUSY (GUT) theories 
with see-saw mechanism of neutrino mass generation \cite{BorzMas86}.
If the SUSY breaking occurs via soft terms with 
universal boundary conditions at a scale $M_X$ above the RH Majorana
neutrino mass scale $M_R$, $M_X > M_R$, as in
the so-called \emph{minimal supergravity}  (mSUGRA) scenario
\cite{Chamseddine:1982jx},
the renormalisation group effects transmit the LFV from 
the neutrino mixing at $M_X$ 
to the effective mass terms of the scalar leptons at $M_R$,
generating new LFV corrections to the flavour-diagonal mass terms. 
For slepton masses of a few hundred GeV, 
the LFV mass corrections at $M_R$ are typically of the order of a few
GeV and thus are much larger than the light neutrino masses $m_j$ . 
As a consequence (and in contrast to the non-supersymmetric case), 
the LFV scalar lepton mixing at $M_R$ generates
additional contributions to the amplitudes of the LFV decays 
and reactions which are not suppressed by the small values 
of neutrino masses. As a result, the LFV processes can
proceed with rates and cross sections which are within the sensitivity 
of presently operating and future planned experiments 
\cite{BorzMas86,Hisano96}
(see also, e.g.,~\cite{Casas:2001sr,Ellis:2002eh,Petcov:2003zb,PS06,Raidal:2008jk}
and the references quoted therein).

  In the following discussion we will assume 
the commonly employed mSUGRA SUSY breaking scenario
\cite{Chamseddine:1982jx}. In this scenario the  
flavour is assumed to be exactly conserved at the GUT scale, 
$M_X\approx 2\times 10^{16}$ GeV, by the soft 
SUSY breaking terms. More specifically, 
it is assumed that at the scale $M_X$
the slepton mass matrices are diagonal in flavour 
and universal, the trilinear couplings 
are proportional to the neutrino and charged lepton Yukawa 
couplings $Y_\nu$ and $Y_e$, respectively, 
and the gaugino masses 
% as well are assumed to 
have a common value:
%%%%%%%%%%%%%%%%%%%%%%%
\ba 
(m^2_{\tilde L} )_{ij} &= &(m^2_{\tilde e})_{ij}= (m^2_{\tilde
\nu})_{ij}=\delta_{ij}m_0^2 \nn\\
% \tilde m^2_{H_d}&=& \tilde m^2_{H_u}=m_0^2\nn\\
A_\nu&=& Y_\nu a_0 m_0, \quad A_e = Y_e a_0 m_0\nn\\
M_{\tilde B} & = &M_{\tilde W} = M_{\tilde g}= m_{1/2}.
\ea 
%%%%%%%%%%%%%%%%%%%%%%%%%%%%%%%%%%%%%%%%%%%%%%%%%%%%%%%%%%%%%%%
%
Hence the parameter space of interest of mSUGRA is determined by:
%%%%%%%%%%%%%%%%%%%%%%%%%%%%%%%%%%%%%%%%%%%%%%%%%%%%%%%%%%%%%%%
\be 
m_{1/2},~ m_0,~ A_0,~ tg\beta\,,
% ~ sgn(\mu).
\ee
%%%%%%%%%%%%%%%%%%%%%%%%%%%%%%%%%%%%%%%%%%%%%%%%%%%%%%%%%%%%%%%
%
$tg\beta$ being  the ration of the vacuum expectation values of 
the Higgs fields $H_u$ and $H_d$.

\begin{figure}[t!]
   \begin{center}
 \subfigure
   {\includegraphics[width=7.5cm]{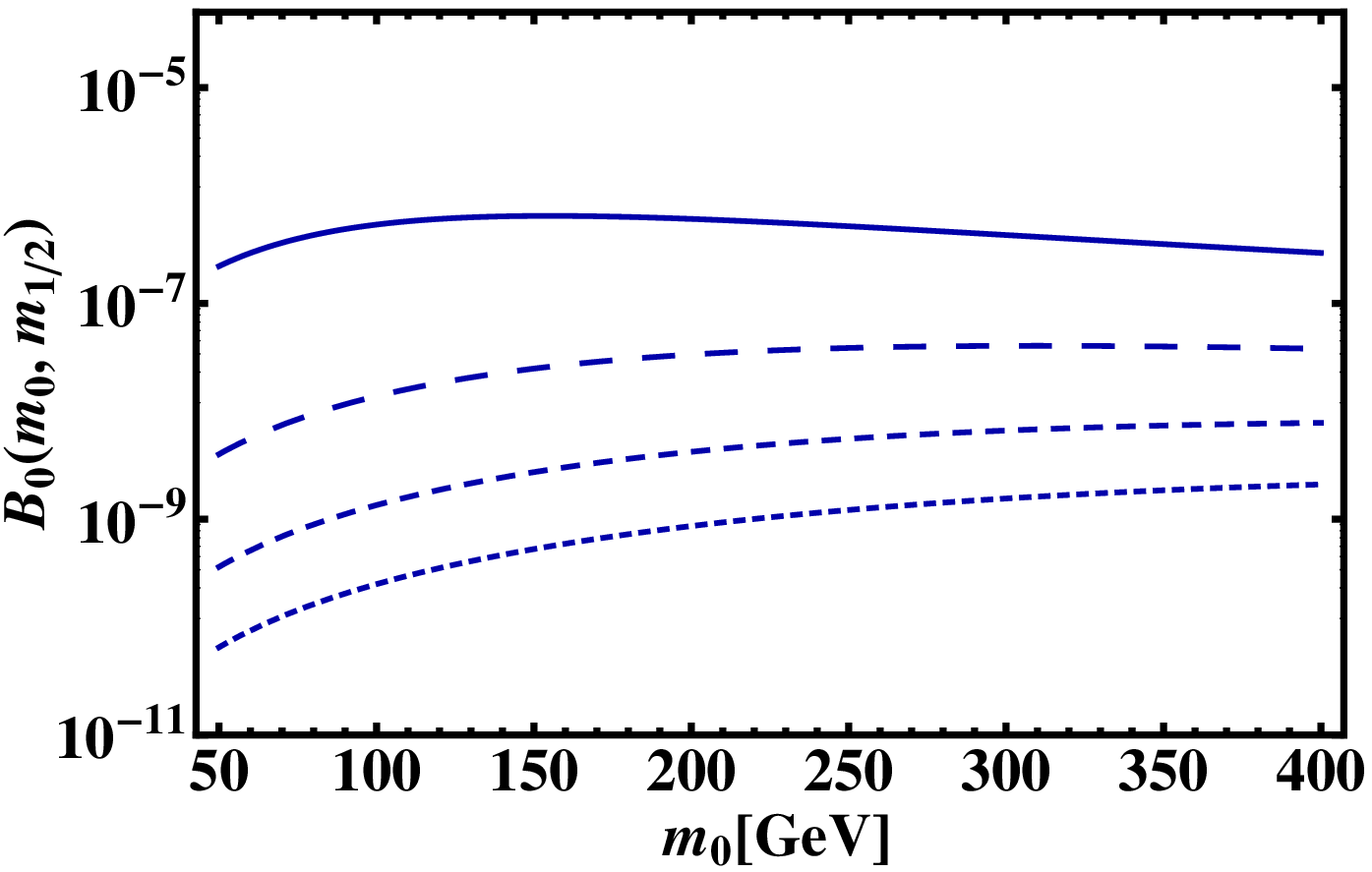}}
 \vspace{5mm}
 \subfigure
   {\includegraphics[width=7.5cm]{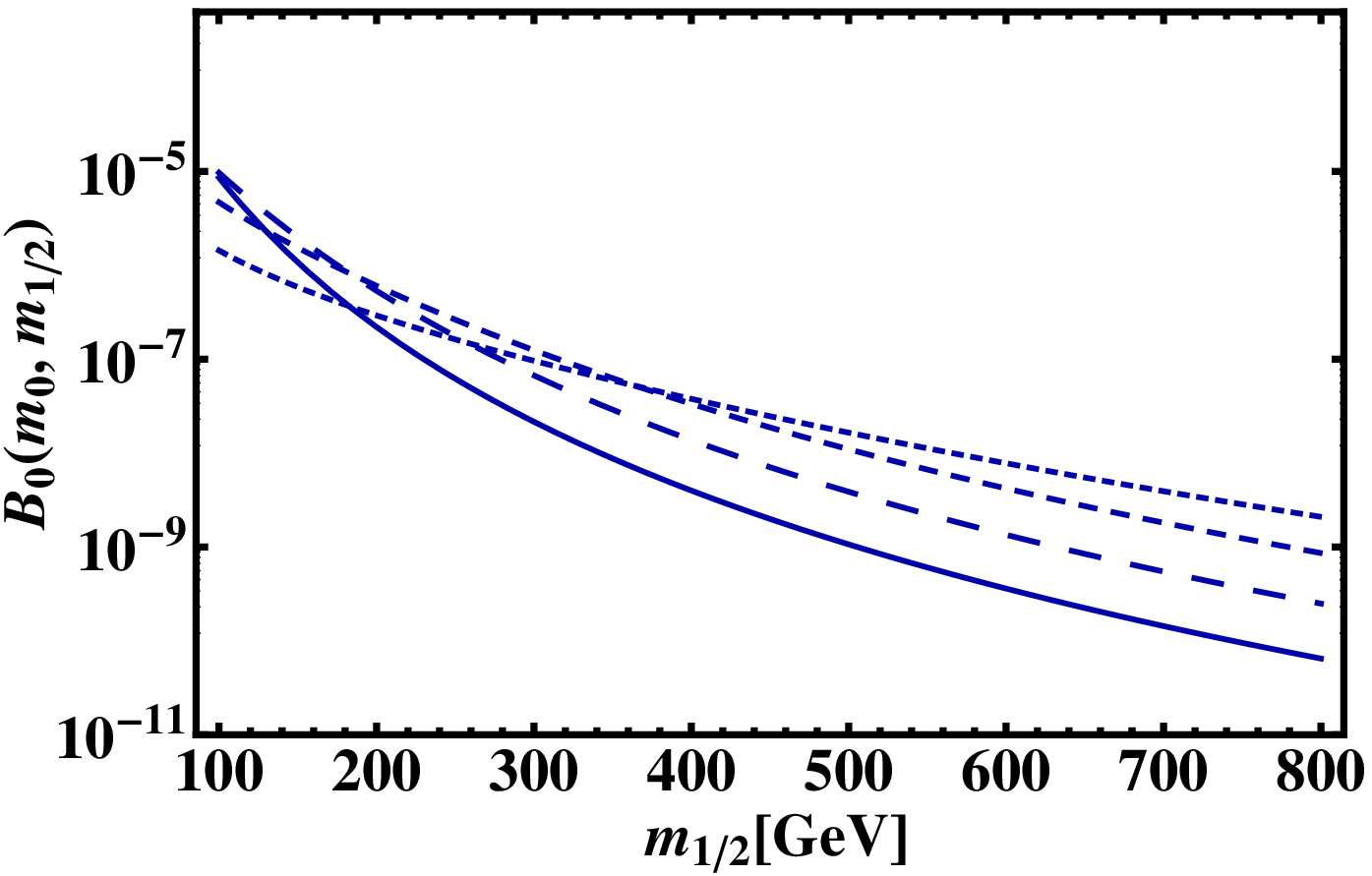}}
     \end{center}
   \caption{The scaling factor $B_0(m_0,m_{1/2})$ as function of $m_0$
 (left panel) and $m_{1/2}$ (right panel) for  $A_0= 7m_0$. 
Left panel: the  values of
$m_{1/2}$ used are 200 GeV (solid line),  400 GeV (long-dashed
line),  600 GeV (medium long-dashed line), 800 GeV (short-dashed
line).  Right panel: the values of $m_0$ used are 50 GeV
(solid line),  100 GeV (long-dashed line),  200 GeV (medium
long-dashed line), 400 GeV (short-dashed line).  
\label{fig:B0}}
\end{figure}

  In the leading-log approximation, the branching ratios 
of the LFV processes $\ell_i \rightarrow\ell_j +\gamma $  
($m_{l_i} > m_{l_j}$ ) is given by \cite{Hisano96} 
(see also, e.g., \cite{Casas:2001sr,Ellis:2002eh,Petcov:2003zb,Raidal:2008jk}):
%%%%%%%%%%%%%%%%%%%%%%%%%%%%%%%%%%%%%%%%%%%%%%%%%%%%%%%%%%%%%%%
\be 
B(\ell_i \rightarrow\ell_j +\gamma )\cong \frac{\Gamma(\ell_i
\rightarrow \ell_j \nu \bar \nu)}{\Gamma_{tot}(\ell_i)} B_0(m_0,m_{1/2})
 \left| \sum_k
(Y_\nu)_{ik} \ln \frac{M_X}{M_k}\ln \frac{M_X}{M_k} (Y^\dagger_\nu)_{kj}\right|^2
\tan^2 \beta.
\label{BR}
\ee
%%%%%%%%%%%%%%%%%%%%%%%%%%%%%%%%%%%%%%%%%%%%%%%%%%%%%%%%%%%%%%%
%
Here $G_F$ is the Fermi constant, $\alpha_{em}\approx1/137$ 
is the fine structure constant, $M_k$ is the mass of the 
heavy Majorana neutrino $N_k$, $Y_\nu$ is the matrix of neutrino 
Yukawa couplings in the basis in which 
the Majorana mass matrix of the RH neutrinos and the matrix 
of charged lepton Yukawa couplings are diagonal, and 
\cite{Nakamura:2010zzi}
$\Gamma(\mu\rightarrow e \nu \bar \nu)/\Gamma_{tot}(\mu)\approx 1$,
$\Gamma(\tau \rightarrow e \nu \bar \nu)/\Gamma_{tot}(\tau)\approx 0.1785$,
$\Gamma(\tau \rightarrow \mu \nu \bar \nu)/\Gamma_{tot}(\tau)\approx 0.1736$.
The scaling function $B_0(m_0,m_{1/2})$ 
contains the dependence on the SUSY breaking parameters:
%%%%%%%%%%%%%%%%%%%%%%%%%%%%%%%%%%%%%%%%%%%%%%%%%%%%%%%%%%%%%%%
 \be 
B_0(m_0,m_{1/2})=\frac{\alpha_{em}^3}{G_F^2 m_S^8} 
\left|\frac{(3+a_0^2)m_0^2}{8\pi^2}\right|^2\,.
\ee
%%%%%%%%%%%%%%%%%%%%%%%%%%%%%%%%%%%%%%%%%%%%%%%%%%%%%%%%%%%%%%%
%
The effective SUSY mass parameter $m_S$ that appears 
in equation (\ref{BR}) can be approximated by \cite{Petcov:2003zb}:
%%%%%%%%%%%%%%%%%%%%%%%%%%%%%%%%%%%%%%%%%%%%%%%%%%%%%%%%%%%%%%%
\be 
m_S^8\approx 0.5 m^2_0 m^2_{1/2} (m^2_0 + 0.6 m^2_{1/2})^2
\ee
%%%%%%%%%%%%%%%%%%%%%%%%%%%%%%%%%%%%%%%%%%%%%%%%%%%%%%%%%%%%%%%
%
This analytic expression was shown to reproduce the exact RG results for 
$B(\ell_i \rightarrow\ell_j +\gamma )$ with high precision.
In Fig. \ref{fig:B0} we illustrate the dependence of 
the scaling function $B_0(m_0,m_{1/2})$
on the parameter $m_0$ ($m_{1/2}$) for 
four values of $m_{1/2}$  ($m_0$).
%%%%%%%%%%%%%%%%%%%%%%%%%%%%%%%%%%%%%%%%
%%%%%%%%%%%%%%%%%%%%%%%%%%%%%%

%%%%%%%%%%%%%%%%%%%%%%%%%%%
%
\subsection{Predictions of the $SU(5)\times T^\prime$ Model}
%
%%%%%%%%%%%%%%%%%%%%%%%%%

 As we have seen, in the  $SU(5)\times T^\prime$ model of flavour 
we are considering the three heavy Majorana neutrinos $N_j$ 
are, to leading order, degenerate in mass: 
$M_j = s_0\Lambda \equiv M > 0$, $j=1,2,3$.
The higher order corrections to the masses $M_j$ 
lead to exceedingly small effects in the 
$\ell_i \rightarrow\ell_j +\gamma $ decay 
rates and we will neglect them.
In this case the branching ratios 
$B(\ell_i \rightarrow\ell_j +\gamma )$ of interest, 
as it follows from eq. (\ref{BR}), depend on the quantity
$|(Y_\nu Y^\dagger_\nu)_{ij}|^2$. 

 As can be shown, in the model under study, 
the matrix of neutrino Yukawa couplings 
$Y_{\nu}$ entering into the expressions for 
$B(\ell_i \rightarrow\ell_j +\gamma )$
is related to the Dirac neutrino mass matrix 
$M_{D}$, eq. (\ref{MD}), as follows:
%%%%%%%%%%%%%%%%%%%%%%%%%%
\be 
Y_{\nu} = \frac{1}{v_u}\,U^\dagger_e \, M_D\, S\,
\label{Ynu1}
\ee
%%%%%%%%%%%%%%%%%%%%%%%%%%%%
%
where $U_e$ is the matrix diagonalising $M_eM^{\dagger}_e$,
$M_e$ being the charged lepton mass matrix, 
and $S$ is determined in eqs. (\ref{MRRdiag}) and (\ref{S}).
In the basis in which the RH neutrino Majorana mass matrix and the
matrix of charged lepton Yukawa couplings are diagonal, 
the Majorana mass term for the LH flavour neutrinos, generated 
by the see-saw mechanism, is given by:
%%%%%%%%%%%%%%%%%%%%%%
\be
M_{\nu} = -\,v^2_u \, Y_{\nu} D_{N}^{-1} Y_{\nu}^{T} = 
U^{\prime}\, D_{\nu}\, (U^{\prime})^T \;,~
\label{MnuYnu}
\ee
%%%%%%%%%%%%%%%%%%%%%
%
where $D_{\nu} \equiv diag(m_1,m_2,m_3)$  and 
%%%%%%%%%%%%%%%%%%%%%%%%
\be
U^{\prime} = U^{\dagger}_e U_{TBM} Q\,.
\label{Uprime}
\ee
%%%%%%%%%%%%%%%%%%%%%%%%%%%%%
%
Equation (\ref{MnuYnu}), as is well known, 
allows to express $Y_{\nu}$ in terms 
of $U^{\prime}$, $D_{\nu}$, $D_{N}$ and 
a complex, in general, orthogonal matrix 
\cite{Casas:2001sr} $R$, 
$R^TR=RR^T = {\bf 1}$:
%%%%%%%%%%%%%%%%%%%%%%%%%%%%
\be 
Y_\nu =\frac{1}{v_u}\, U^{\prime}\,\sqrt{D_{\nu}}\,R\,\sqrt{D_N}\,. 
\label{YnuR}
\ee
%%%%%%%%%%%%%%%%%%%%%%
%
From eqs. (\ref{Ynu1}) -  (\ref{YnuR}) and 
(\ref{eq:MDdiag}), we obtain the following 
expression for the matrix $R$: 
%%%%%%%%%%%%%%%%%%%%%%%%
\be
R  = v_u\,(\sqrt{D_{\nu}})^{-1}\,
Q^*\, \,M^{diag}_D\, \,U^T_{TBM}\  S\,(\sqrt{D_N})^{-1}\,.
\label{R}
\ee
%%%%%%%%%%%%%%%%%%%%%%%%%%%
%
Using the explicit form of $Q=diag(1,1,\pm i)$ 
and of $U_{TBM}$ and $S$, 
eqs. (\ref{TBMM}) and (\ref{S}),  
it is not difficult to show that 
$R$ is a real matrix: $R^* = R$.
Taking into account this result and 
recalling that  $D_N = M\,diag(1,1,1)$, we get 
for the quantity of interest  
$Y_\nu Y^\dagger_\nu$:
%%%%%%%%%%%%%%%%%%%%
\be
Y_\nu Y^\dagger_\nu = 
\frac{M}{v^2_u}\,U^\prime\,D_{\nu}\,(U^\prime)^\dagger\,,
\label{YnuYnudag}
\ee
%%%%%%%%%%%%%%%%%%%%%%
%
where the PMNS matrix $U^\prime = U^\dagger_e U_{TBM} Q$, see
eqs. (\ref{PMNSTBM1}), (\ref{PMNSTBM2}), (\ref{PMNSTBM3}) 
and  (\ref{PMNSnumeric}).
Thus, the $\ell_i \rightarrow\ell_j +\gamma$ 
decay branching ratios in the $SU(5)\times T^\prime$ 
model under investigation depend on the mass of the 
heavy Majorana neutrinos through the factor 
$(M(\ln M_X/M_k)/v_u)^2$, do not depend on the matrix $R$, 
and their ratios are entirely determined 
by the elements of the PMNS matrix and the light 
neutrino masses. More specifically, using 
the general expressions for the elements of
the PMNS matrix and the unitarity of the latter 
we get \cite{PS06}:
%%%%%%%%%%%%%%%%%%%%%%%%%%%%%%%%%%%%%%%%%%
\be 
\left |(Y_{\nu}Y^\dagger_{\nu})_{\mu e}\right | =
\left | U_{\mu j}m_{j}U^*_{e j}\right | 
\simeq \left | \Delta_{21}\, s_{12}c_{23}c_{12}  
% - s^2_{12}s_{23}s_{13}e^{i\delta} \Delta_{21}
+ \Delta_{31}\, s_{23}s_{13}e^{i\delta} \right|\,,
\label{YYmue}
\ee
%%%%%%%%%%%%%%%%%%%%%%%%%%%%%%%%%%%%%%%%%%
%
\be
\left |(Y_{\nu}\,Y^{\dagger}_{\nu})_{\tau e}\right | =
\left | U_{\tau j}\,m_{j}\,U^*_{e j}\right | 
\simeq \left |-\,\Delta_{21}\, s_{12}c_{12}s_{23}
% - s^2_{12}c_{23}s_{13}e^{i\delta}\Delta_{21}
+ \Delta_{31}\,c_{23}s_{13}e^{i\delta} \right|\,,
\label{YYtaue}
\ee
%%%%%%%%%%%%%%%%%%%%%%%%%%%%%%%%%%%%%%%
\be
\left | (Y_{\nu}\,Y^{\dagger}_{\nu})_{\tau \mu}\right| =
\left | U_{\tau j}\,m_{j}\,U^*_{\mu j} \right | 
\simeq \left| 
% -\,\left (\,c^2_{12} c_{23}s_{23}-\, s_{12}c_{12}s_{13}
% (s^2_{23}e^{-i\delta} - c^2_{23}e^{i\delta})\right )\Delta_{21} + 
\Delta_{31}\, c_{23}s_{23} \right|\,,
\label{YYtaumu}
\ee
%%%%%%%%%%%%%%%%%%%%%%%%%%%%%%%%%%%%%
%
where
%%%%%%%%%%%%%%%%%%%%%%%%%%%%%%%
\be
\Delta_{ij}= \frac{M_R}{v_u^2}\frac{\Delta
m^2_{ij}}{m_i+m_j}\,.
\ee
%%%%%%%%%%%%%%%%%%%%%%%%%%%%%%%
%
In eqs. (\ref{YYmue}) - (\ref{YYtaumu}) 
we have used the approximation 
$c_{13} \cong 1$ and have neglected  
$c^2_{12} \Delta_{21}$, 
$s^2_{12} \Delta_{21}$ and 
higher order terms $\propto \Delta_{21}$ 
with respect to $\Delta_{31}$.

We define the ``double'' ratios of the branching 
ratios $B(\ell_i \rightarrow\ell_j +\gamma )$ as follows:
%%%%%%%%%%%%%%%%%%%%%%%%%%%%%%%%%%%%%%%
\be
R(21/31)\equiv \frac{BR(\mu\rightarrow e
\gamma)}{BR(\tau\rightarrow e \gamma)}BR(\tau\rightarrow e
\nu_{\tau} \bar\nu_e ),
\quad R(21/32)\equiv \frac{BR(\mu\rightarrow
e \gamma)}{BR(\tau\rightarrow \mu \gamma)}BR(\tau \rightarrow e
\nu_{\tau} \bar\nu_e )
\label{R1213}
\ee
%%%%%%%%%%%%%%%%%%%%%%%%%%%%%%%%%%
%
The double ratios of interest are given by:
%%%%%%%%%%%%%%%%%%%%%%%%%%%%%%%%%%%%%%%
\be
R(21/31) \cong \frac{|(Y_{\nu}Y^{\dagger}_{\nu})_{\mu e}|^2}
{|(Y_{\nu}Y^{\dagger}_{\nu})_{\tau e}|^2}\,,\\
\quad R(21/32)\cong \frac{|(Y_{\nu}Y^{\dagger}_{\nu})_{\mu e}|^2}
{|(Y_{\nu}Y^{\dagger}_{\nu})_{\tau \mu}|^2}\,.
\label{R1213}
\ee
%%%%%%%%%%%%%%%%%%%%%%%%%%%%%%%%%%
%

%%%%%%%%%%%%%%%%%%%%%%%%%%%%%%%%%%%%%%%%%%%%%%%%%%%
\subsection{Numerical Results}
%%%%%%%%%%%%%%%%%%%%%%%%%%%%%%%%%%%%%%%%%%%%%%%%%%%

\begin{figure}[tbh!]
   \begin{center}
\subfigure
   {\includegraphics[width= 0.325\textwidth]{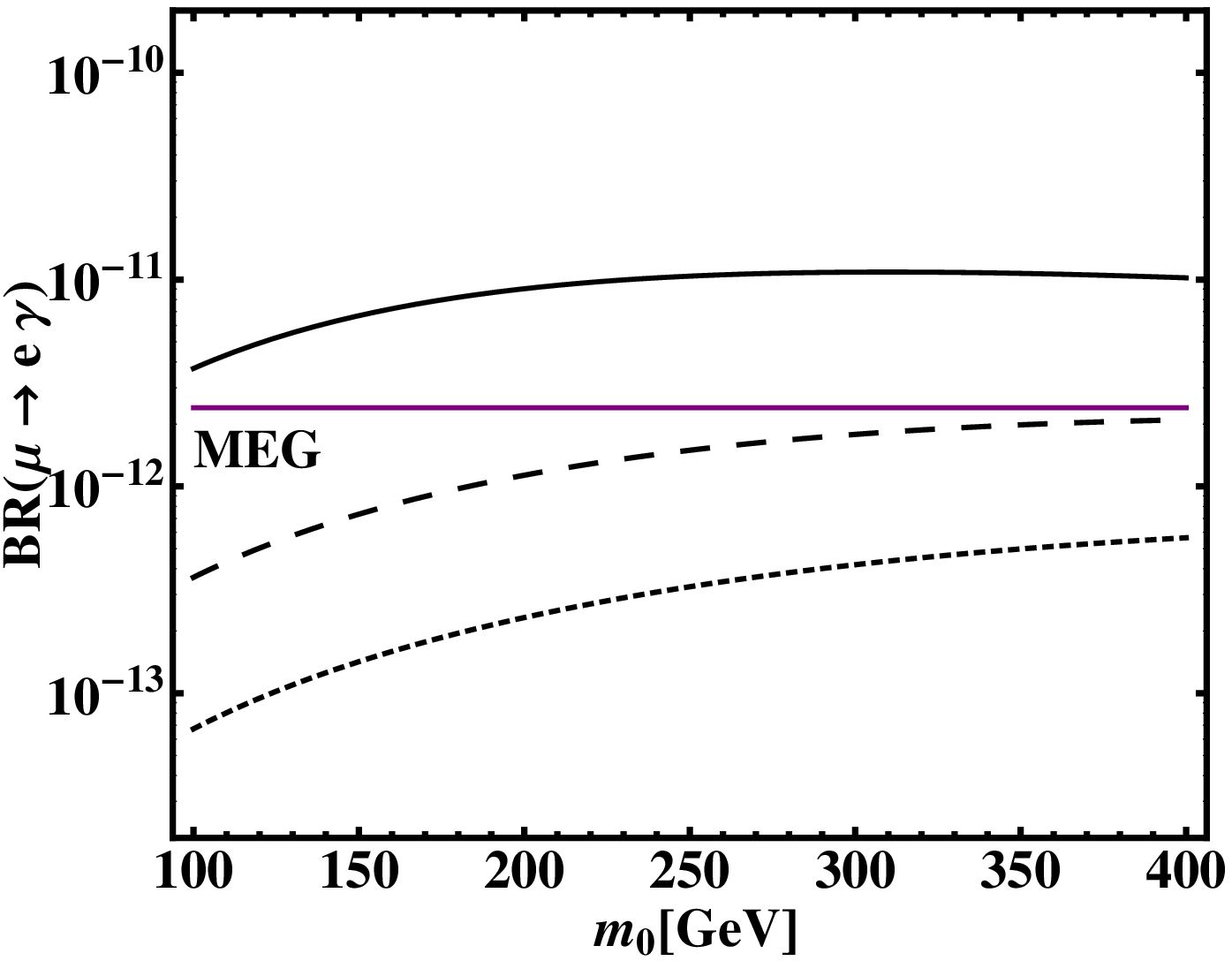}}
% \vspace{5mm}
\subfigure
   {\includegraphics[width=0.325\textwidth]{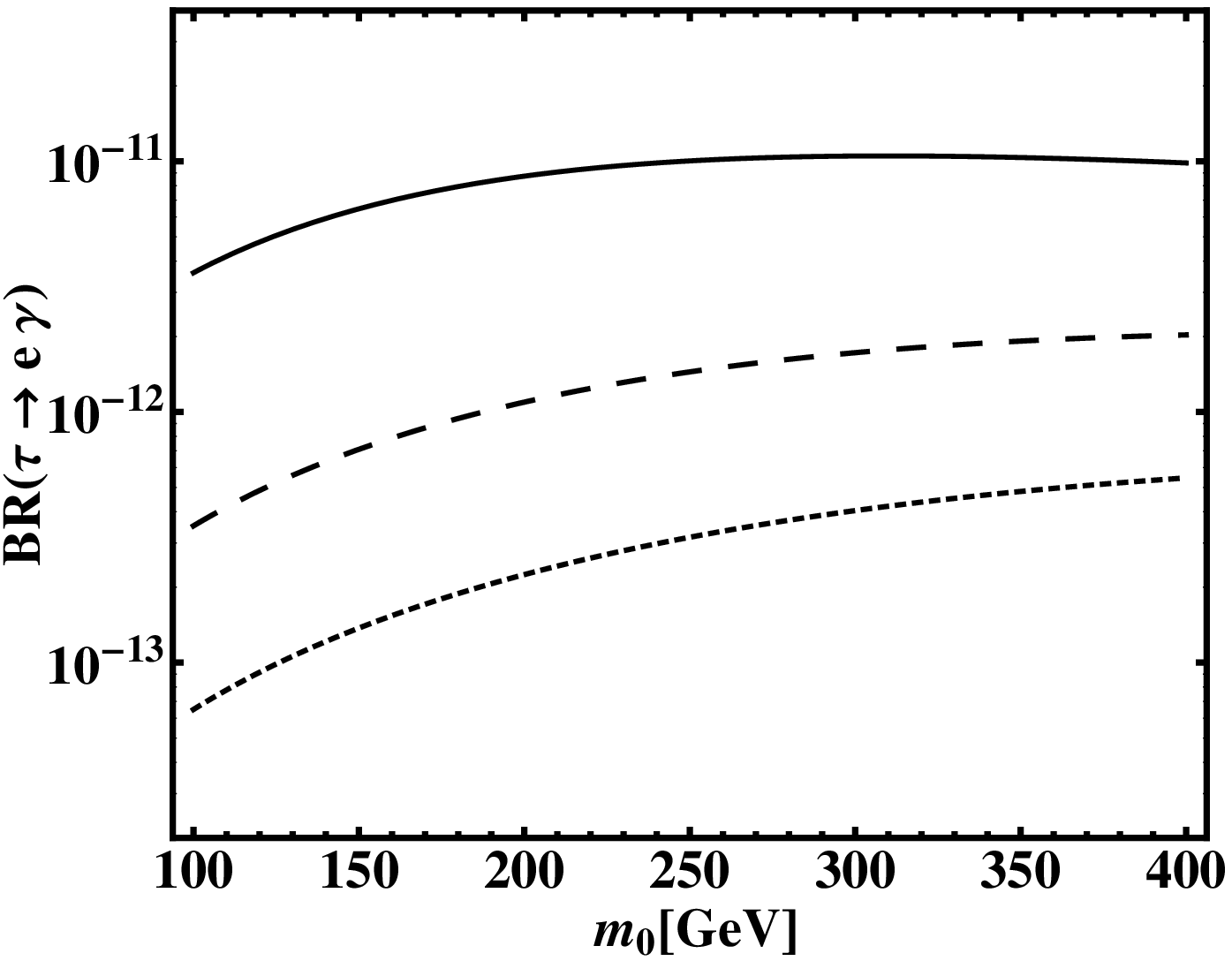}}
 \vspace{5mm}
 \subfigure
   {\includegraphics[width=0.325\textwidth]{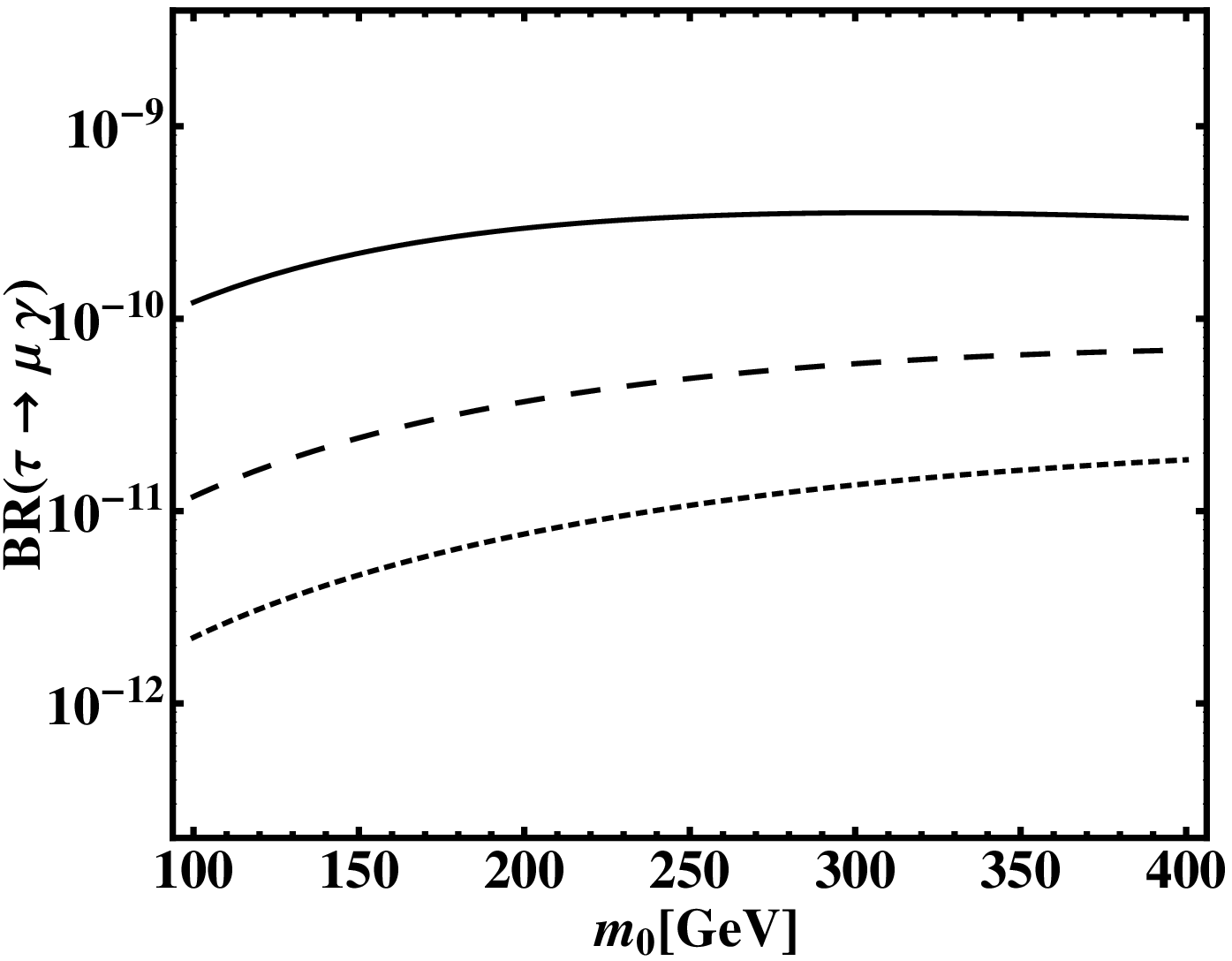}}
     \end{center}
   \caption{The branching ratios  BR($\mu\rightarrow e+ \gamma$) (left),
BR($\tau \rightarrow e + \gamma$) (center) and BR($\tau \rightarrow
\mu + \gamma$) (right) as function of $m_0$ for 
$\Delta m^2_{21} = 7.58\times 10^{-5}~{\rm eV^2}$,
$r=0.032$, $\delta=5\pi/4$, $\sin\theta_{13}$= 0.058, 
$\tan\beta=28$, $A_0$= 7$m_0$,
and three values of $m_{1/2}$: 400 GeV (solid line), 600 GeV
(dashed line), 800 GeV (dotted line). The horizontal
line corresponds to the MEG bound BR($\mu \rightarrow e + \gamma$)$ <
2.4 \times 10^{-12} $. See text for details.
\label{fig:BRm0}}
\end{figure}

\begin{figure}[tbh!]
   \begin{center}
\subfigure
   {\includegraphics[width= 0.325\textwidth]{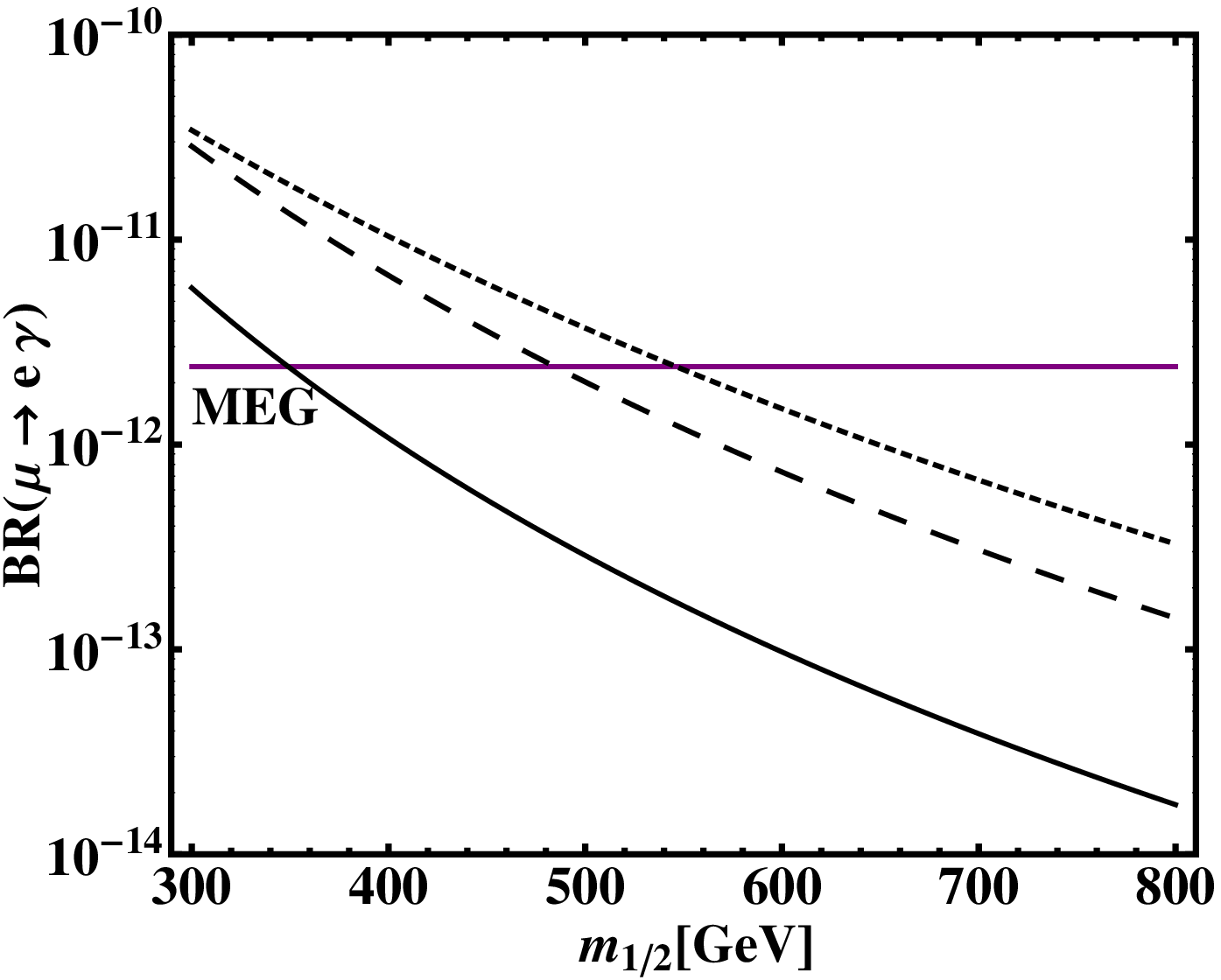}}
% \vspace{5mm}
\subfigure
   {\includegraphics[width=0.325\textwidth]{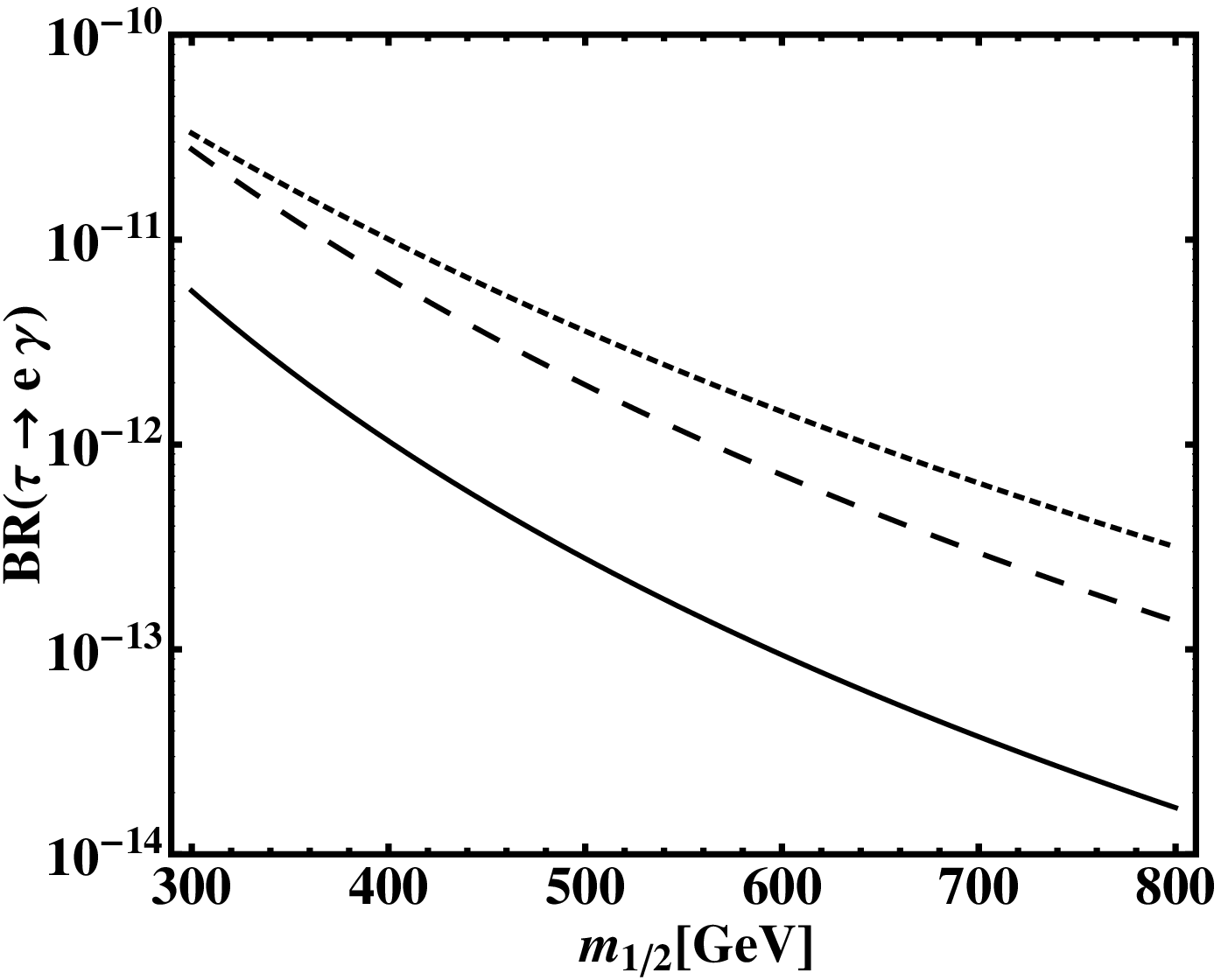}}
 \vspace{5mm}
 \subfigure
   {\includegraphics[width=0.325\textwidth]{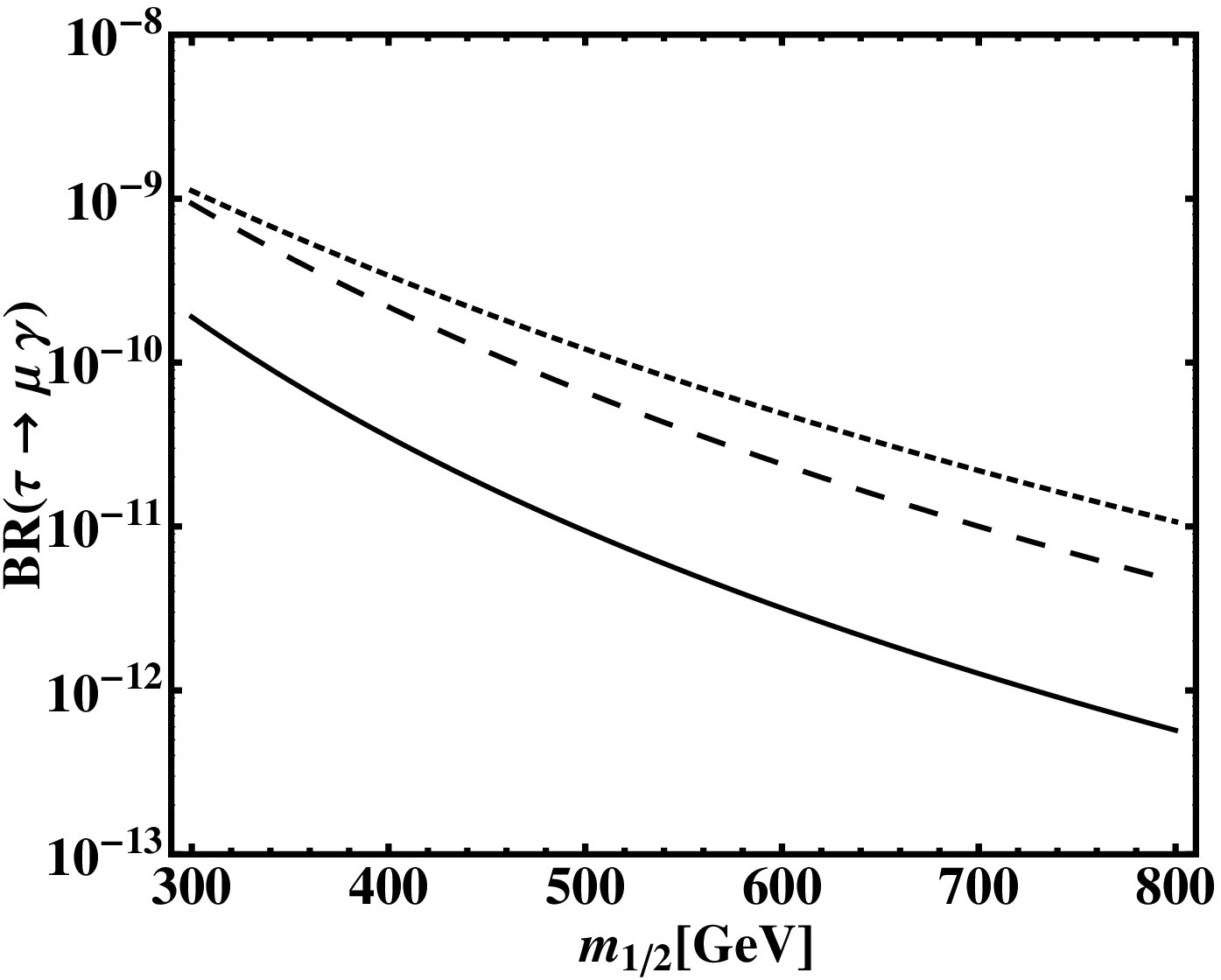}}
     \end{center}
   \caption{The branching ratios  $BR(\mu\rightarrow e + \gamma)$ (left),
$BR(\tau \rightarrow e + \gamma)$ (center) and $BR(\tau \rightarrow
\mu + \gamma)$ (right) as function of $m_{1/2}$ 
for three values of $m_{0}$: 50 GeV (solid line), 150 GeV (dashed
line), 250 GeV (dotted line). The values of the 
other parameters used are the same as those quoted 
in the caption of Fig. \ref{fig:BRm0}.
The horizontal line corresponds to 
the MEG bound BR($\mu \rightarrow e + \gamma$)$ <
2.4 \times 10^{-12} $. 
\label{fig:BRmh}
}
\end{figure}

\begin{figure}[tbh!]
   \begin{center}
\subfigure
   {\includegraphics[width= 0.325\textwidth]{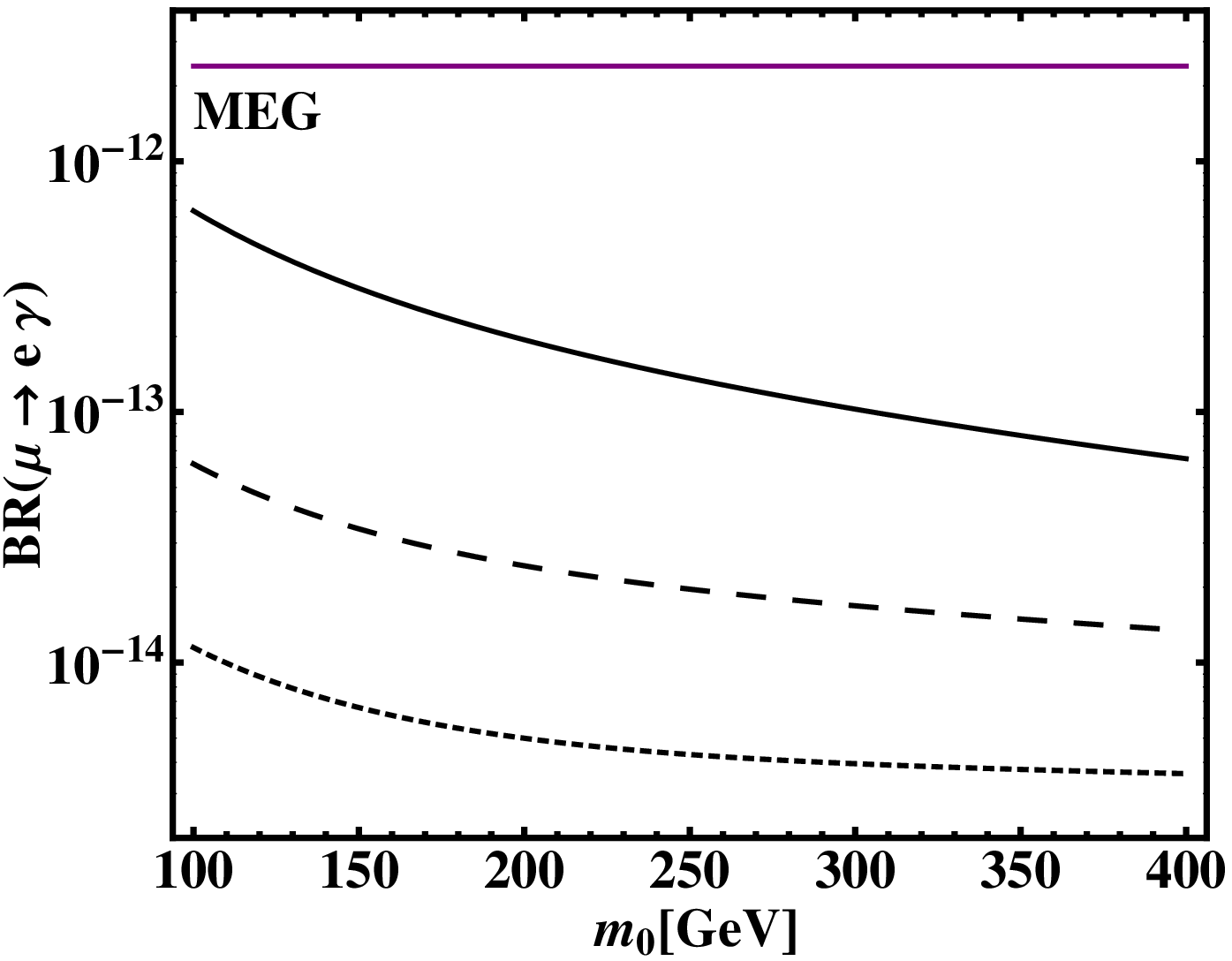}}
% \vspace{5mm}
\subfigure
   {\includegraphics[width=0.325\textwidth]{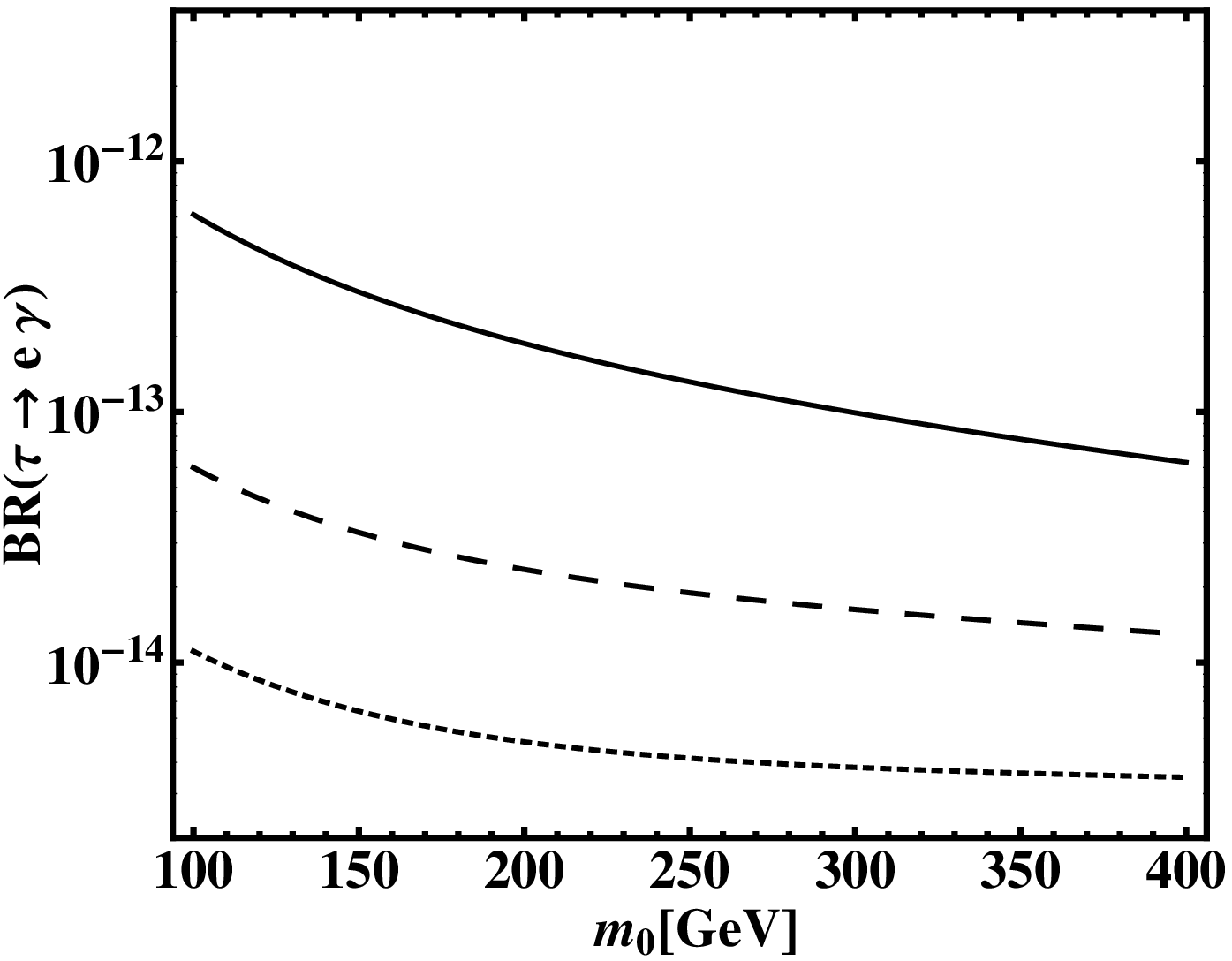}}
 \vspace{5mm}
 \subfigure
   {\includegraphics[width=0.325\textwidth]{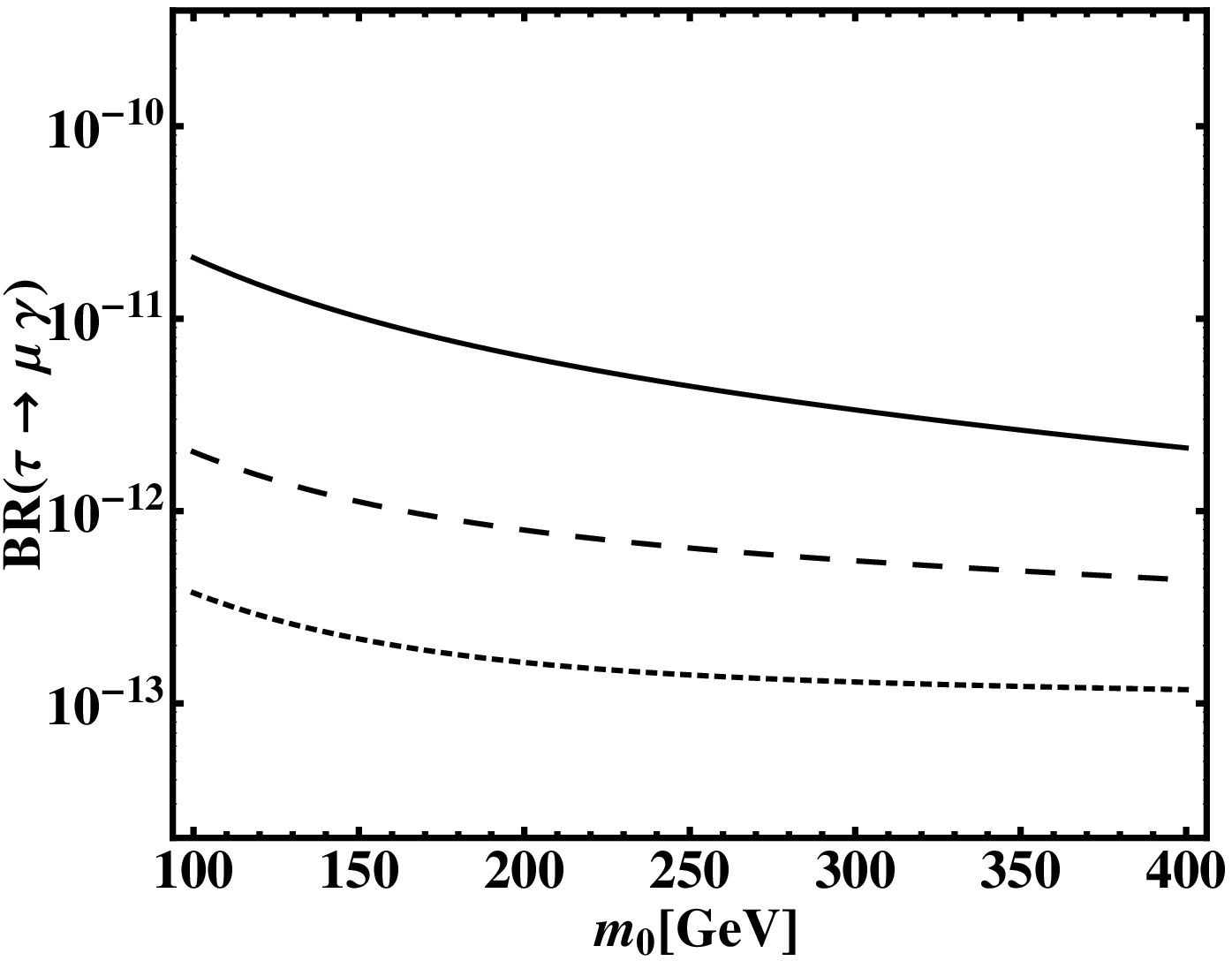}}
     \end{center}
   \caption{The branching ratios  BR($\mu\rightarrow e + \gamma$) (left),
BR($\tau \rightarrow e + \gamma$) (center) and BR($\tau \rightarrow
\mu + \gamma$) (right) as function of $m_0$
for $A_0 = 430$ GeV and three values of 
$m_{1/2}$: 400 GeV (solid line), 600 GeV (dashed line), 800 GeV
(dotted line). The values of the 
other parameters used are the same as those quoted 
in the caption of Fig. \ref{fig:BRm0}.
The horizontal line corresponds to the MEG bound 
$BR(\mu \rightarrow e + \gamma) < 2.4 \times 10^{-12} $. 
\label{fig:BRs1}
}
\end{figure}

\begin{figure}[tbh!]
   \begin{center}
\subfigure
   {\includegraphics[width= 0.325\textwidth]{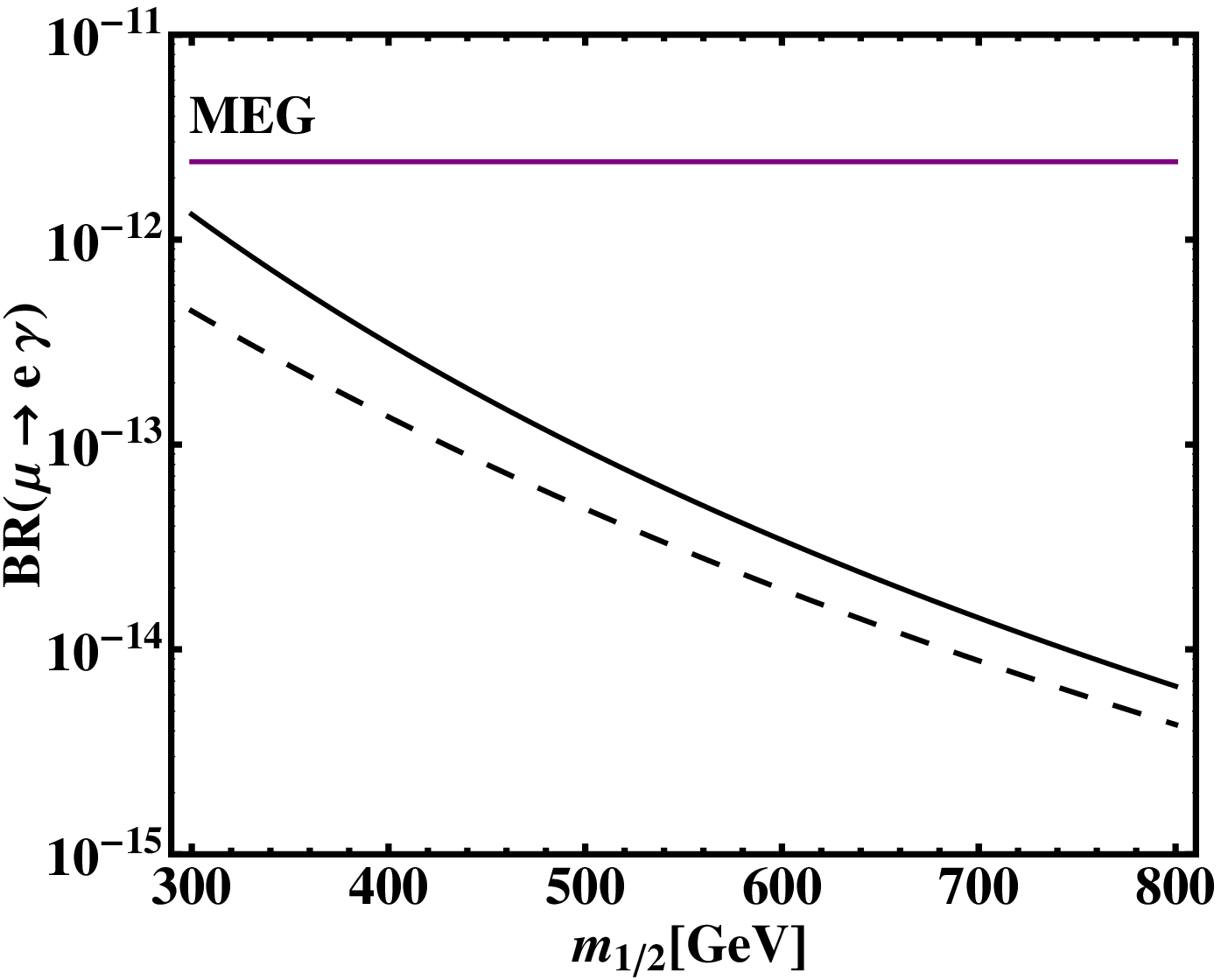}}
% \vspace{5mm}
\subfigure
   {\includegraphics[width=0.325\textwidth]{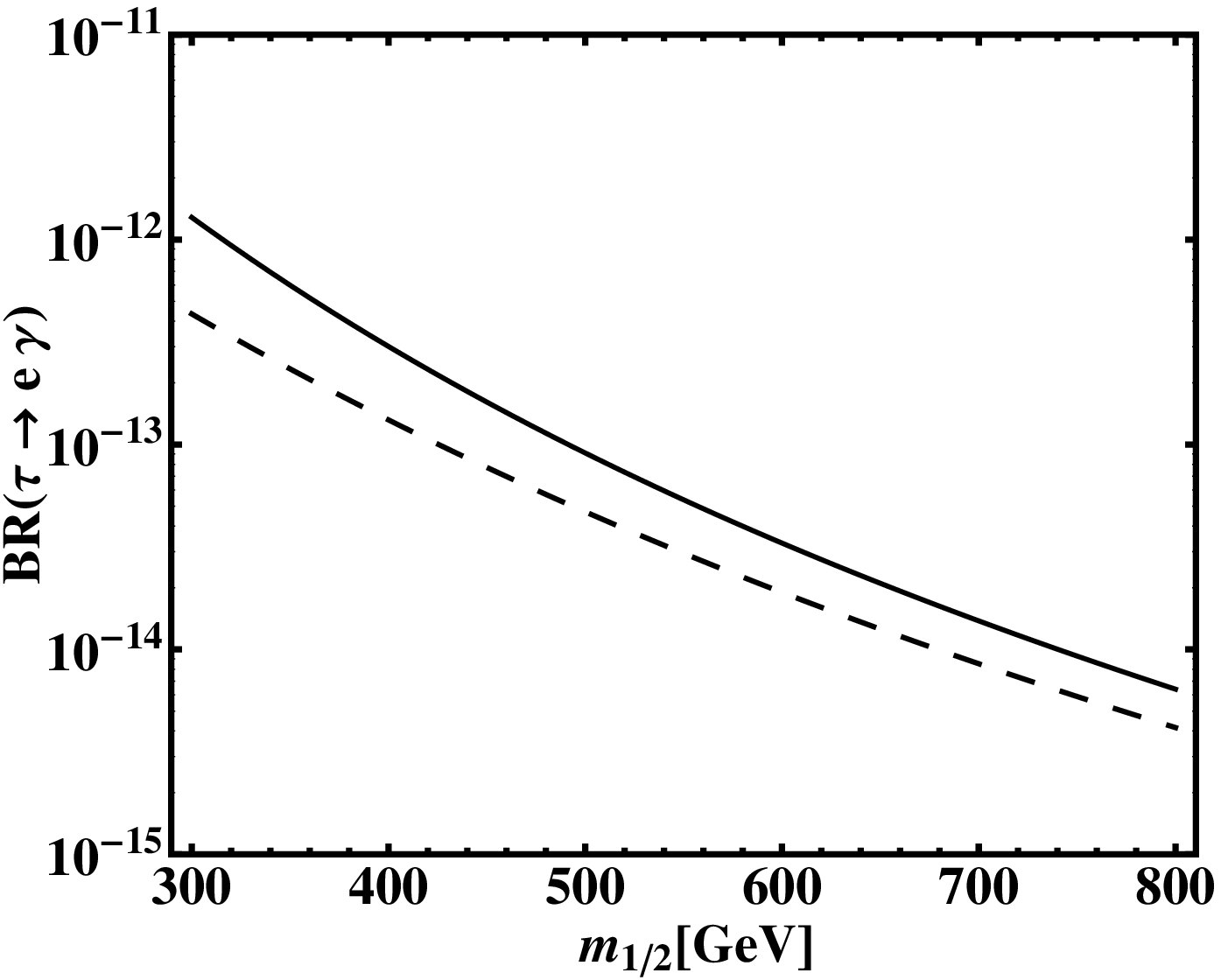}}
 \vspace{5mm}
 \subfigure
   {\includegraphics[width=0.325\textwidth]{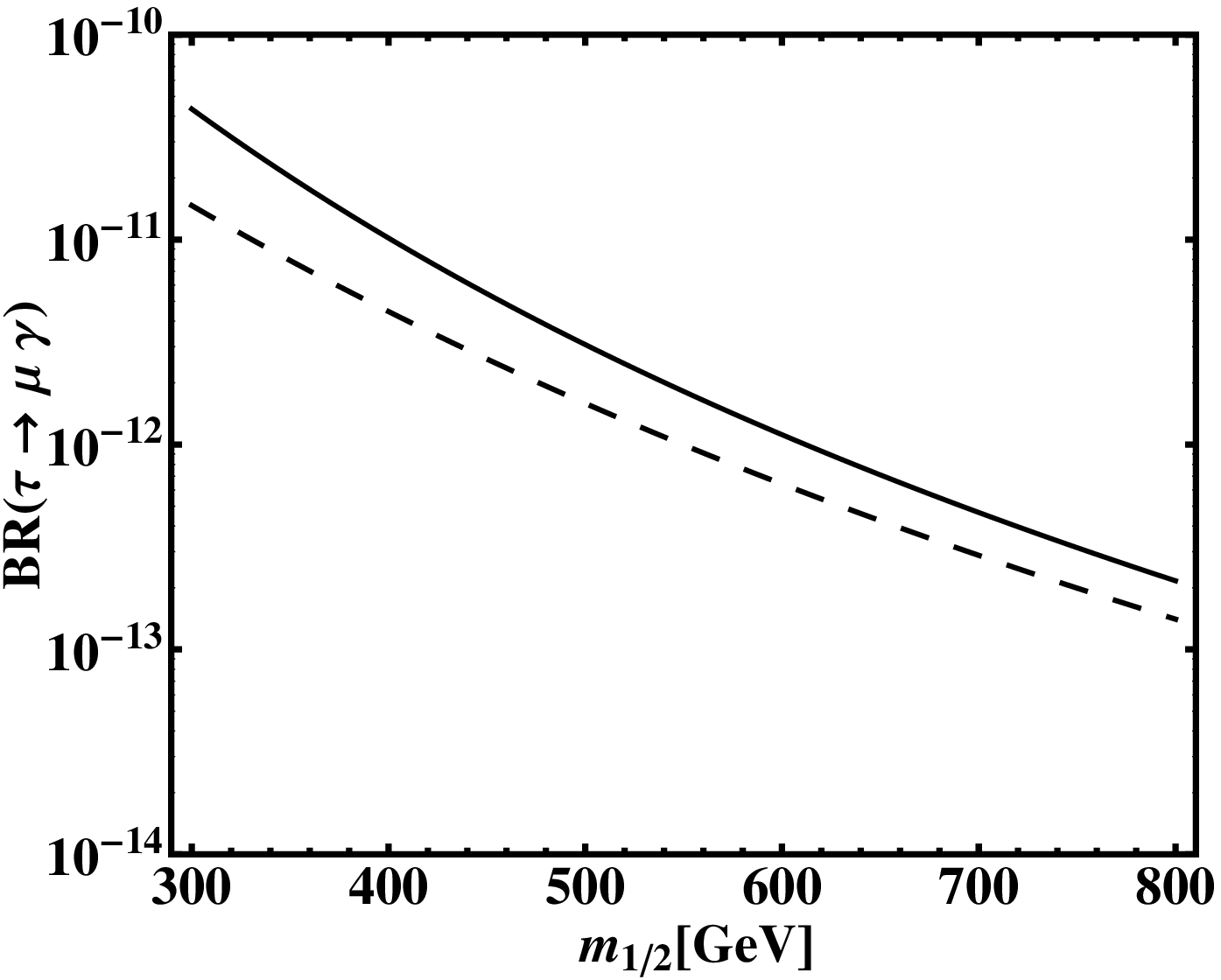}}
     \end{center}
   \caption{The branching ratios  BR($\mu\rightarrow e + \gamma$) (left),
BR($\tau \rightarrow e + \gamma$) (center) and BR($\tau \rightarrow
\mu + \gamma$) (right) as function of $m_{1/2}$
for  $A_0 = 430$ GeV and $m_{0} = 150$ GeV (solid line), 250 GeV (dashed line).
The values of the other parameters used are the same 
as those quoted in the caption of Fig. \ref{fig:BRm0}.
The horizontal dashed line corresponds to the 
MEG bound BR($\mu \rightarrow e+\gamma)<
2.4 \times 10^{-12} $. 
\label{fig:BRs2}
}
\end{figure}

 In Figs. \ref{fig:BRm0} - \ref{fig:BRs2}  we present results on 
the branching ratios $BR(\mu\rightarrow e + \gamma)$,
$BR(\tau\rightarrow e + \gamma)$ and 
$BR(\tau\rightarrow \mu+ \gamma)$,
predicted by the $SU(5)\times T^\prime$ model. 
The branching ratios are 
calculated using eq. (\ref{BR}) with $M_k$ and $M_X$ 
set to $M_k = M = 10^{12}$ GeV and $M_X =2\times 10^{16}$ GeV. 
The values of the mSUGRA parameters $\tan\beta$ and $A_0$ 
are chosen from the intervals 
$\tan\beta= 3 \div 50$, $A_0= 0 \div 7m_0$, 
which are compatible with the constaints 
obtained by the ATLAS \cite{Aad:2011hh, Zhuang:2011qq} 
and CMS \cite{Khachatryan:2011tk} experiments 
at LHC. The values of the other two 
relevant mSUGRA parameters, $m_0$ and  $m_{1/2}$, 
are chosen from intervals favored by the global data 
analysis performed  \cite{Buchmueller:2011aa}. 
The data set used in this analysis includes in addition to 
the results of the ATLAS and the CMS  experiments, 
the data on the muon $(g-2)$, on the precision 
electroweak observables, on $B$-physics observables, 
astrophysical data on the cold dark matter density,
as well as the limits from the direct 
searches for Higgs boson and sparticles at LEP.
Based on the results obtained in 
\cite{Buchmueller:2011aa},  $m_0$ and  $m_{1/2}$
are chosen from, or to vary in, the intervals 
 $50~{\rm GeV} \leq m_0 \leq  400$ GeV and 
$300~{\rm GeV}\leq m_{1/2} \leq 800$ GeV.
For the values of the lepton 
mixing angles and the Dirac phase $\delta$ we use 
those predicted (to leading order) by the 
$SU(5)\times T^\prime$ model (unless otherwise stated):
$\sin^2\theta_{23} = 0.5$, $\sin^2\theta_{12} = 0.299$, 
$\sin\theta_{13} = 0.058$, $\delta = 5\pi/4$.
The neutrino masses used as input 
are obtained using the best fit values of 
$\Delta m^2_{21} = 7.58\times 10^{-5}~{\rm eV^2}$ 
and $r= 0.032$. 
%%%%%%%%%%%%%%%%%%%%%%%%%%%%%%%%%%%%%%%%
%%%%%%%%%%%%%%%%%%%%%%%%%%%%%%

%%%%%%%%%%%%%%%%%%%%%%%%%%%%%%%%%%%%%%%%
%%%%%%%%%%%%%%%%%%%%%%%%%%%%%%
%

 As the Figs. \ref{fig:BRm0} - \ref{fig:BRs2} show, 
for the values of $M = 10^{12}$ GeV and $\tan\beta = 28$ 
considered and $m_0$ from the interval (50 - 400) GeV, 
the predictions for the branching ratios
$BR(\mu\rightarrow e \gamma)$,
$BR(\tau \rightarrow e \gamma)$ and 
$BR(\tau \rightarrow\mu \gamma)$
are very sensitive to the 
values of $m_{1/2}$:  
when the latter increases from 
300 GeV to  800 GeV, the branching ratios 
decrease by approximately 2 orders of magnitude.
For the fixed value of $A_0 = 430$ GeV, 
we have $BR(\mu\rightarrow e \gamma)\gtap 10^{-13}$
provided  
%%%%%%%%%%%%%%%%%%%%%%%%%%%%%%%%%%%%%%%%
%%%%%%%%%%%%%%%%%%%%%%%%%%%%%%

%%%%%%%%%%%%%%%%%%%%%%%%%%%%%%%%%%%%%%%%
%%%%%%%%%%%%%%%%%%%%%%%%%%%%%%
%
\noindent
$m_0 \ltap 300$ GeV and 
 $m_{1/2} \ltap 400$ GeV 
(Figs. \ref{fig:BRs1} and \ref{fig:BRs2}).
The ranges of values of $m_0$  and 
 $m_{1/2}$, for which  
$10^{-13} \ltap BR(\mu\rightarrow e \gamma) < 2.4\times 10^{-12}$,
is very different in the case when $A_0 \propto m_0$.
For $A_0 = 7m_0$ ( Figs. \ref{fig:BRm0} - \ref{fig:BRmh}), 
for instance, $BR(\mu\rightarrow e \gamma)$ satisfies the MEG 
upper limit and is in 
%%%%%%%%%%%%%%%%%%%%%%%%%%%%%%%%%%%%%%%%
\begin{figure}[tbh!]
   \begin{center}
\subfigure
   {\includegraphics[width= 0.45\textwidth]{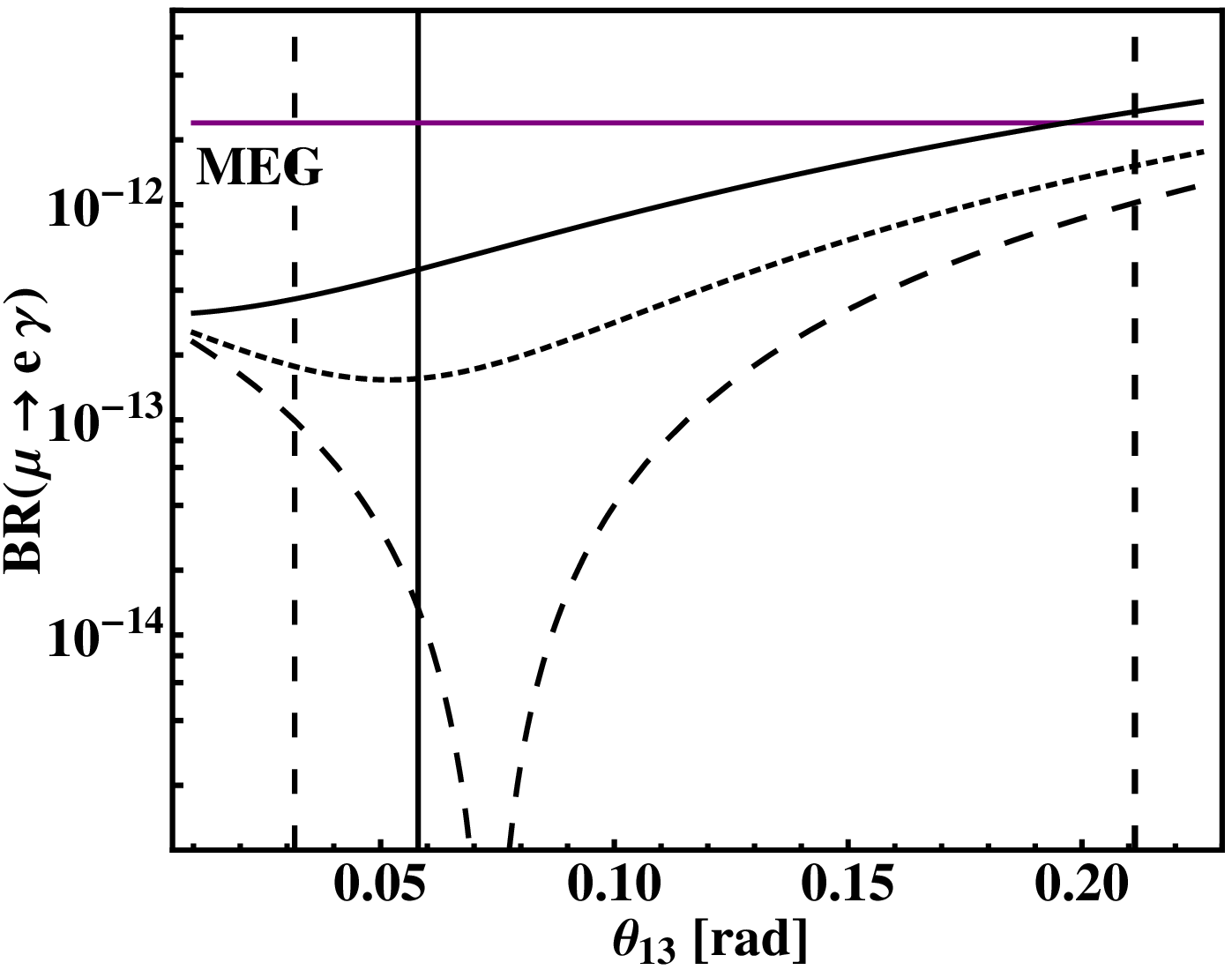}}
\hspace{6mm}
\subfigure
   {\includegraphics[width=0.45\textwidth]{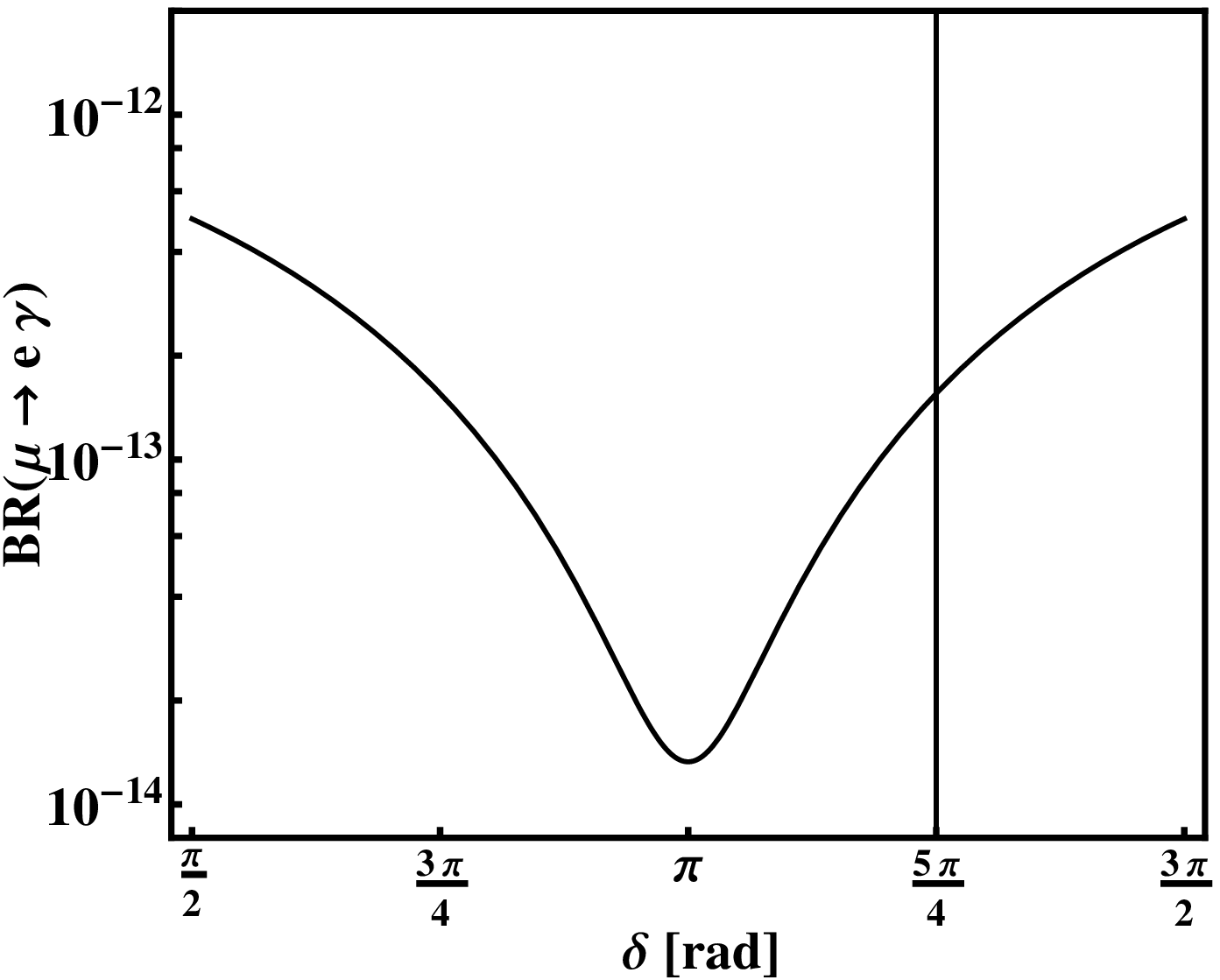}}
     \end{center}
   \caption{The branching ratio BR($\mu \rightarrow e + \gamma$)
as function of $\theta_{13}$ (left panel) 
and of the Dirac phase $\delta$ (right panel)
for  $r=0.032$, $m_{0}= 230$ GeV, 
$m_{1/2}=400$ GeV, $A_0=430$ GeV and $\tan\beta=28$. 
In the left panel, $BR(\mu \rightarrow e + \gamma)$ is
plotted for three values of $\delta$: $\pi/2$ (solid line) , 
$\pi$ (dashed line), $ 5\pi/4$ (dotted line).
The horizontal line corresponds to the MEG 
bound $BR(\mu \rightarrow e+\gamma) < 2.4 \times 10^{-12}$. 
The results for $BR(\mu \rightarrow e+\gamma)$ 
shown in the right panel are obtained   
for $\sin\theta_{13}=0.058$. 
The solid vertical line in the right panel 
corresponding to $\delta= 5 \pi/4$. 
\label{fig:BRmuegamma} 
}
\end{figure}
%%%%%%%%%%%%%%%%%%%%%%%%%%%%%%
%
\noindent the range of sensitivity 
of the MEG experiment for $m_0$ lying in the interval 
 $m_0 \cong (100- 300)$ GeV if 
 $600~{\rm GeV} \ltap m_{1/2} \ltap 800$ GeV. 
For the values of the parameters used 
to obtain  Figs. \ref{fig:BRm0} - \ref{fig:BRs2}, 
we find that  
$BR(\tau \rightarrow\mu + \gamma) \ltap 10^{-9}$
and $BR(\tau \rightarrow e + \gamma) \ltap 3\times 10^{-11}$.

The $\mu \rightarrow e + \gamma$ decay
branching ratio $BR(\mu \rightarrow e + \gamma)$
can exhibit very strong dependence on the value of 
the angle $\theta_{13}$ if the Dirac 
phase $\delta \cong \pi$, and on the Dirac phase $\delta$ 
in the case of
$\sin\theta_{13}=0.058$.
This is illustrated in Fig. \ref{fig:BRmuegamma} and can be 
easily understood on the basis of the analytic 
expression for $|(Y_{\nu}Y^\dagger_{\nu})_{\mu e}|$
 given in eq. (\ref{YYmue}). 
As Fig.  \ref{fig:BRmuegamma} shows,
for the values of $\delta = 5\pi/4$ and 
$\sin\theta_{13}=0.058$, predicted by the model, 
$BR(\mu \rightarrow e \gamma)$ is somewhat smaller 
than the maximal value it can have as a function 
of  $\delta$ and $\sin\theta_{13}$.

\begin{figure}[tbh!]
   \begin{center}
\subfigure
   {\includegraphics[width= 0.325\textwidth]{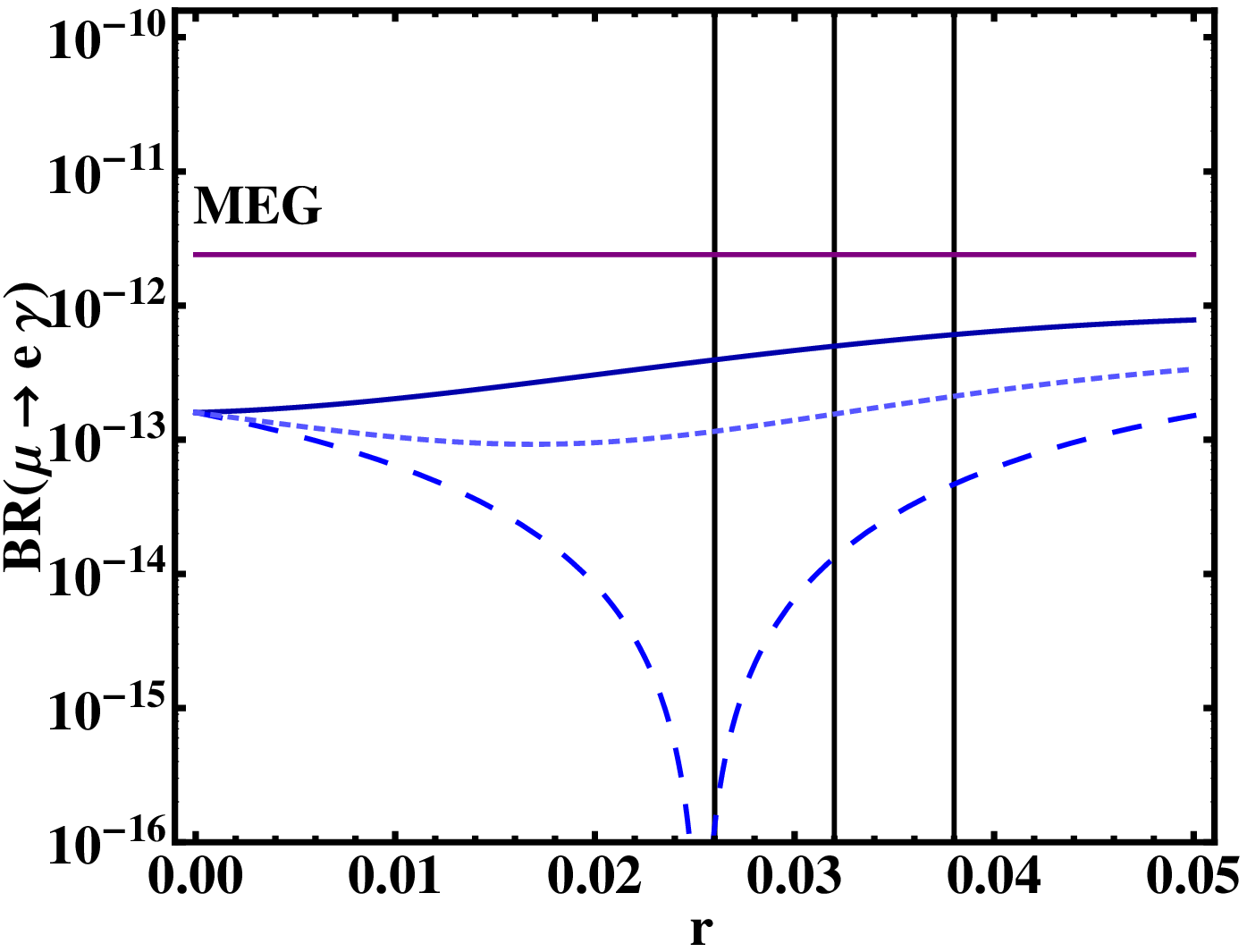}}
% \vspace{5mm}
\subfigure
   {\includegraphics[width=0.325\textwidth]{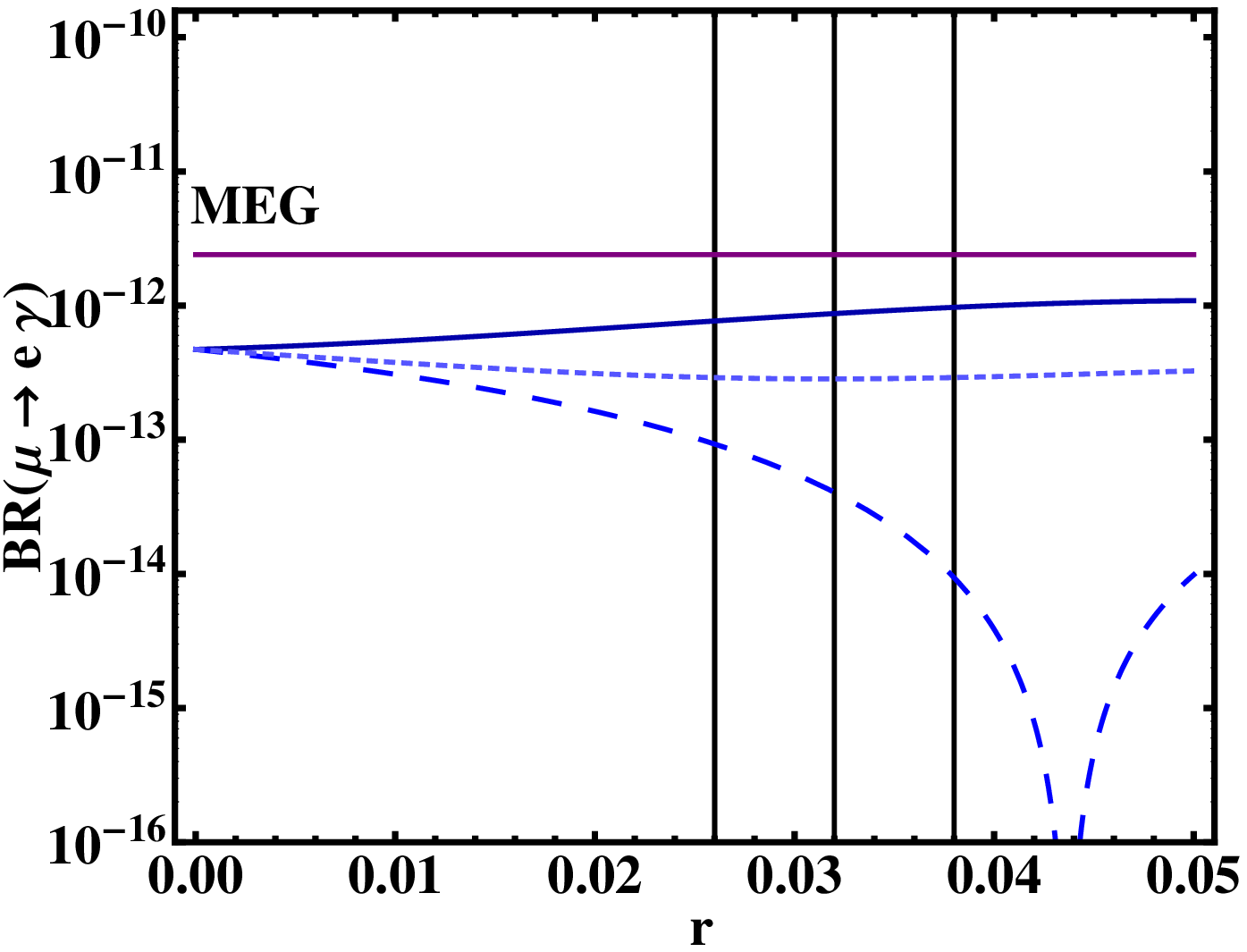}}
 \vspace{5mm}
 \subfigure
   {\includegraphics[width=0.325\textwidth]{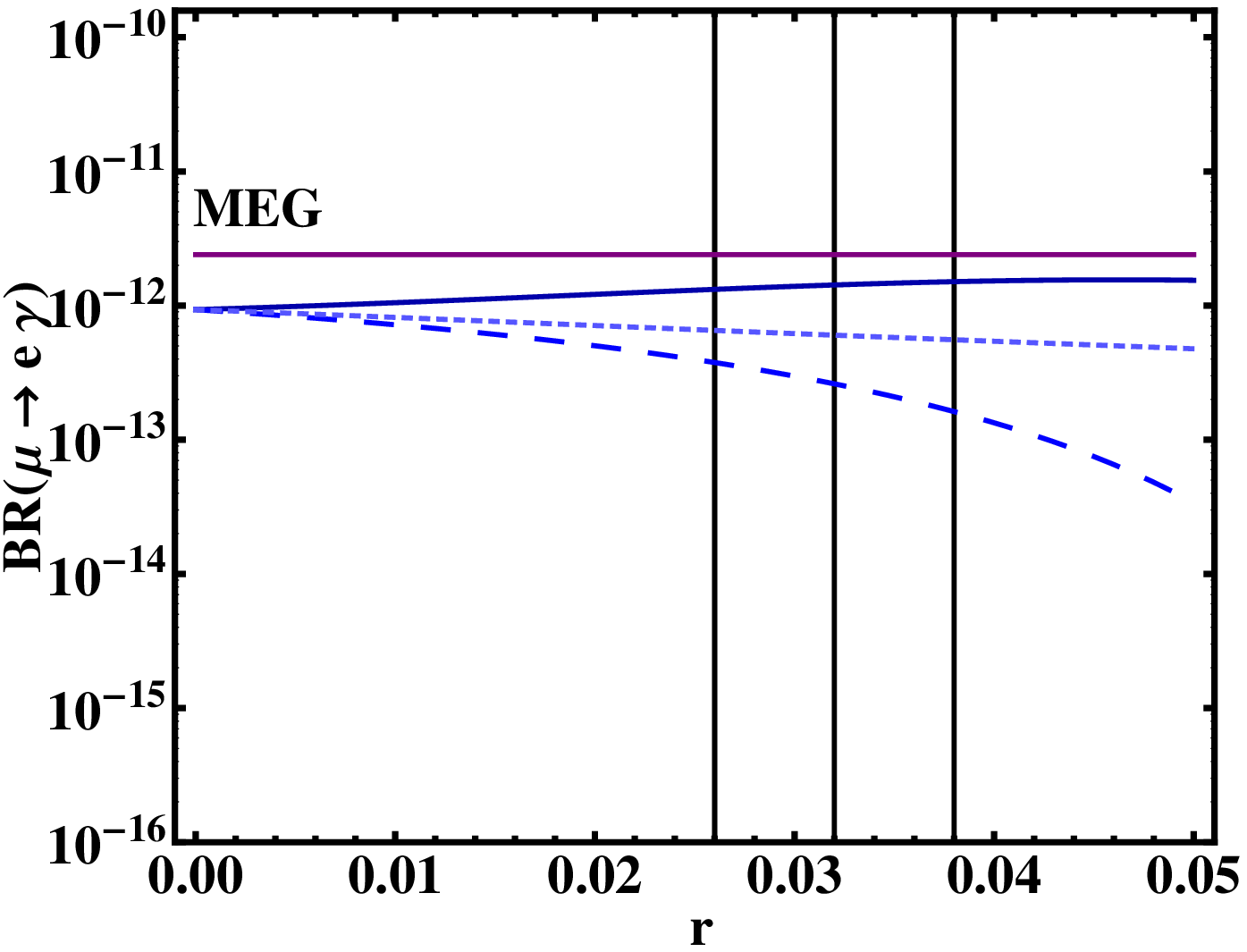}}
     \end{center}
   \caption{The dependence of the branching ratio 
BR($\mu \rightarrow e + \gamma$)
on the parameter $r$ for 
$m_{0}= 230$ GeV, $m_{1/2}= 400$ GeV, 
$A_0=430$ GeV and  $\tan\beta=28$.
The left, middle and right panels 
are obtained for $\sin^2\theta_{13} = 3.4 \times 10^{-3}$ 
\cite{Chen:2009gf}, 0.01 and 0.02, respectively. 
The values of $\delta$ used are  $\pi/2$ (solid line), 
$\pi$ (dashed line) and $5\pi/4$ (dotted line). The vertical lines
correspond respectively to $r=$0.026, 0.032, 0.038. The horizontal
line shows the MEG bound $BR(\mu \rightarrow e+ \gamma) <
2.4 \times 10^{-12}$. See text for details.  
\label{fig:BRrmuegamma}
}
\end{figure}

 We have studied also  
the dependence of the $\mu\rightarrow e + \gamma$ decay 
branching ratio $BR(\mu\rightarrow e + \gamma)$ 
on the parameter $r$. The results of this study are 
illustrated in Fig. \ref{fig:BRrmuegamma}
for three values of $\sin^2\theta_{13}$, 
$\sin^2\theta_{13} = 3.4 \times 10^{-3}, 
0.01,0.02$, and three values of $\delta$, 
$\delta = \pi/2,\pi,5\pi/4$. For illustrative purposes 
the ratio $r$ is varied in the interval $0\leq r\leq 0.05$, 
which is wider than the current $3\sigma$ range of allowed 
values of $r$, $0.026 \leq r \leq 0.038$. Figure 
 \ref{fig:BRrmuegamma} exhibits  in a different way
the sensitivity of $BR(\mu\rightarrow e + \gamma)$ 
to the values of $\theta_{13}$ and $\delta$.
%%%%%%%%%%%%%%%%%%%%%%%%%%%%%%%%%%%%%%%%
%%%%%%%%%%%%%%%%%%%%%%%%%%%%%%
%

\begin{figure}[t!]
   \begin{center}
 \subfigure
   {\includegraphics[width=7cm]{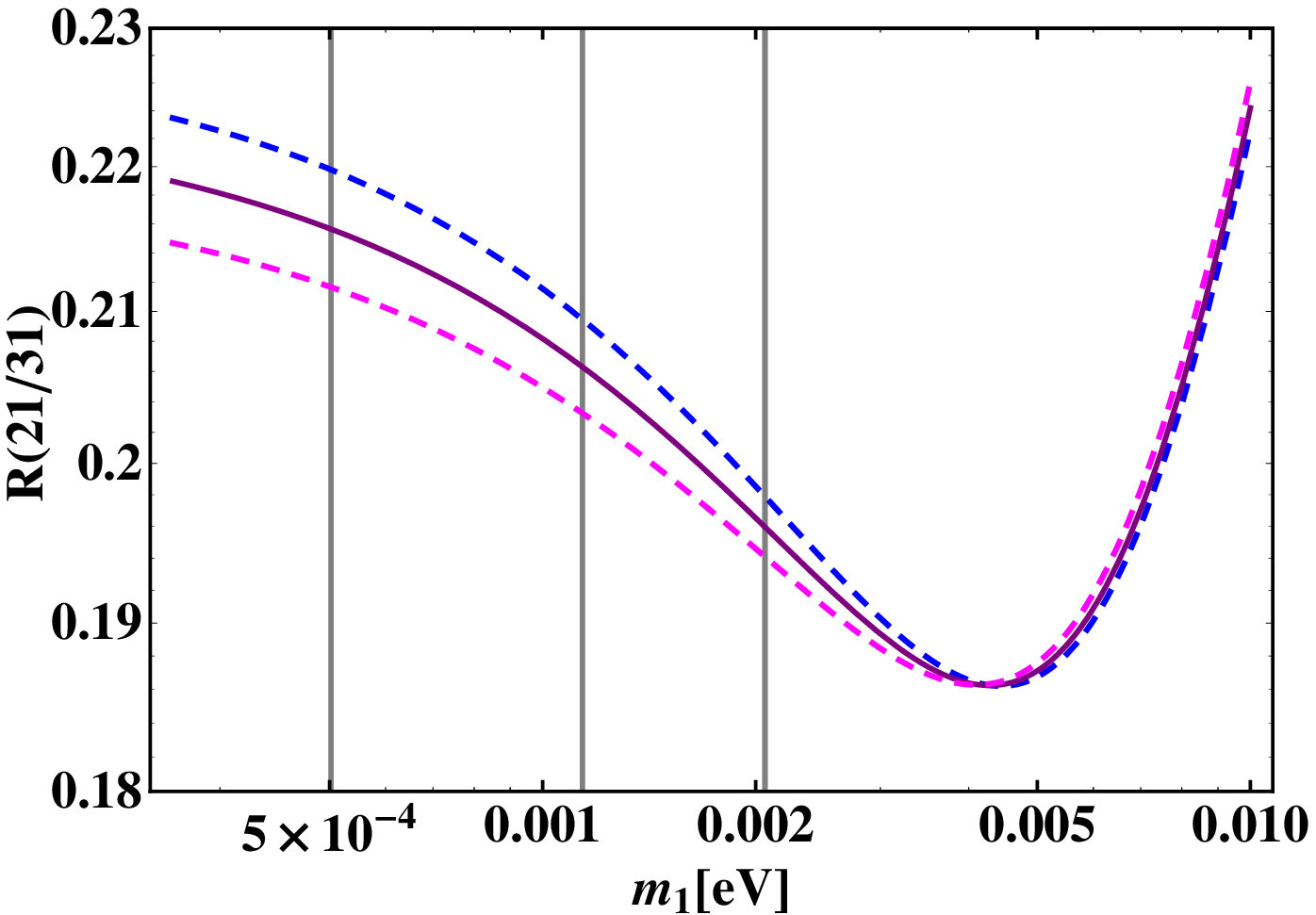}}
 \vspace{5mm}
 \subfigure
   {\includegraphics[width=7cm]{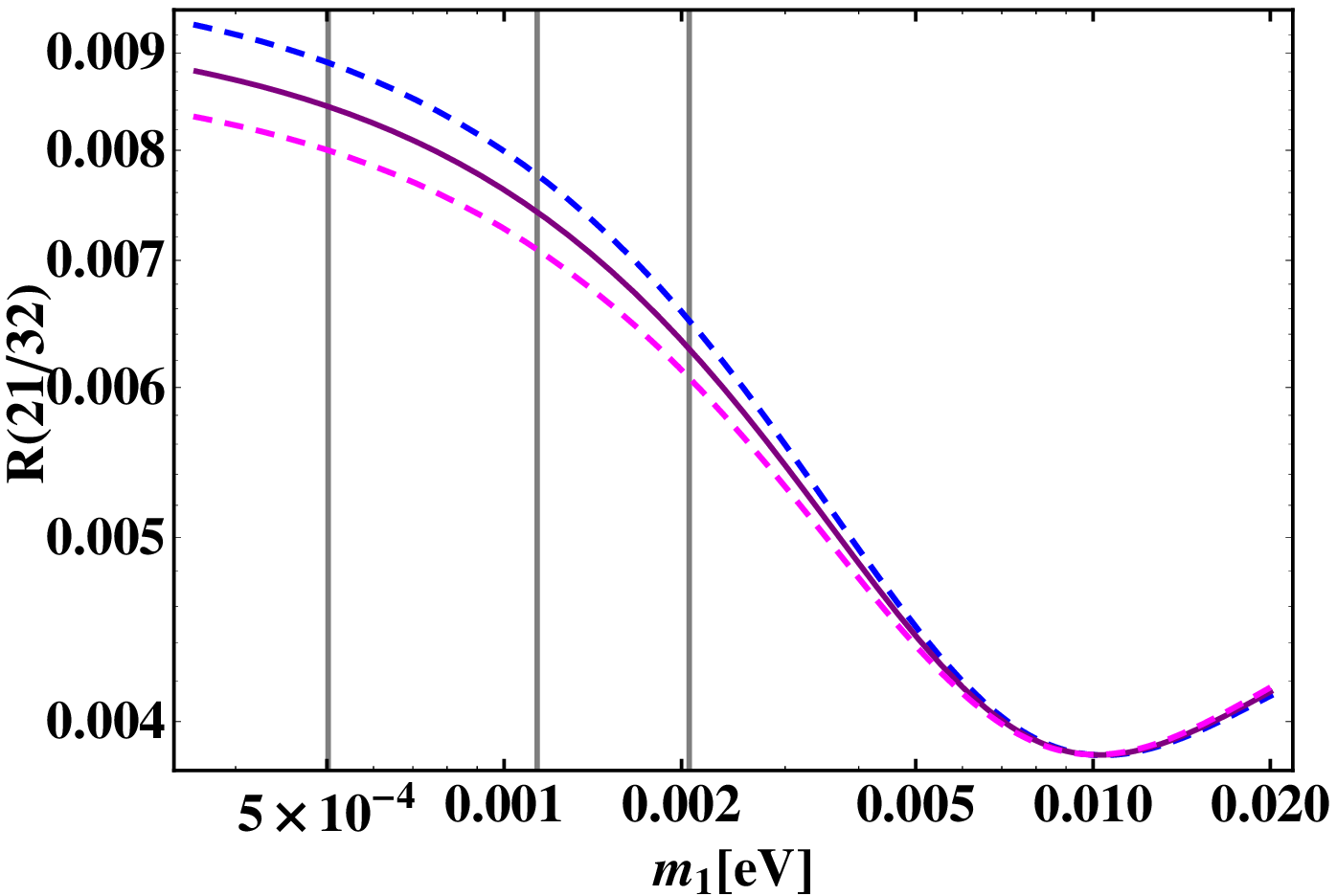}}
     \end{center}
   \caption{The double ratios R(21/31), left panel, 
and R(21/32), right panel,
as functions of the lightest neutrino mass, $m_1$. 
The three vertical lines in each plot are positioned at
$m_1 = 2.06\times 10^{-3}, 1.14\times 10^{-3},~5.0\times 10^{-4}$ eV,
and correspond to $r= 0.026, 0.032, 0.038$ and the best 
fist value of $\Delta m^{2}_{21} = 7.58\times 10^{-5}~{\rm eV^2}$.    
The solid lines are obtained with the best-fit values 
of  $\Delta m^{2}_{21}$ and $\Delta m^{2}_{31}$  given
in Table \ref{tab:tabdata-1106}, while the 
dashed lines indicate the $3\sigma$ allowed ranges.   
\label{fig:Ratios}
}
\end{figure}

In Fig. \ref{fig:Ratios} we present results   
the double ratios $R(21/31)$ and  $R(21/32)$ 
predicted by the model, as a function of the lightest 
neutrino mass $m_1$. We recall that the double ratios 
$R(21/31)$ and  $R(21/32)$ in the model considered 
do not depend on the heavy Majorana neutrino 
mass $M$ and on the mSUGRA parameters: 
they are entirely determined by the values 
of the neutrino masses, the neutrino 
mixing angles and the Dirac CP violating 
phase $\delta$. All these neutrino parameters 
have essentially definite values in 
the $SU(5)\times T^\prime$ model considered.
Thus, the values of the 
double ratios $R(21/31)$ and  $R(21/32)$
are predicted by the model with relatively small 
uncertainties, as is also seen 
in  Fig. \ref{fig:Ratios}: we have numerically
%%%%%%%%%%%%%%%%%%%%%%%%%
\be
R(21/31) \cong 0.21,~~~ R(21/32) \cong 7.4\times 10^{-3}\,.
\label{R213j}
\ee
%%%%%%%%%%%%%%%%%%%%%%%%%%%%%%%
%
These values are one of the characteristic predictions 
of the  $SU(5)\times T^\prime$ model of flavour 
under investigation.
It is interesting to note also that in the model 
with $SU(5)\times T^\prime$ symmetry 
considered, the $\tau \rightarrow \mu + \gamma$
decay branching ratio $BR(\tau\rightarrow \mu + \gamma)/
BR(\tau \rightarrow e \nu_{\tau} \bar\nu_e )$ 
can be bigger than the
$\mu \rightarrow e + \gamma$ decay branching ratio
$BR(\mu\rightarrow e + \gamma)$ by a factor $\sim 10^{2}$.

%%%%%%%%%%%%%%%%%%%%%%%%%%%%%%%%%%%%%%%%
%%%%%%%%%%%%%%%%%%%%%%%%%%%%%%

%%%%%%%%%%%%%%%%%%%%%%%%%%%
%
\section{Conclusions}
%
%%%%%%%%%%%%%%%%%%%%%%%

  In the present article we  have investigate certain 
aspects of the low energy lepton phenomenology  of 
the SUSY  $SU(5)\times T'$ model of flavour which 
was developed in ~\cite{Chen:2007afa,Chen:2009gf}, 
and which allows to describe 
in a unified way the masses and the mixing of  
the quarks and the leptons, neutrinos included, as well as 
the CP violation both in the quark and lepton sectors.
The model includes three right-handed neutrino 
fields which possess a Majorana 
mass term. The light neutrino masses 
are generated by the type I see-saw mechanism and are 
naturally small. The light and heavy neutrinos are 
Majorana particles. The heavy Majorana neutrinos 
$N_k$ are predicted (to leading order) to be 
degenerate in mass: $M_k = M$.
The model is free of discrete gauge 
anomalies.
In addition to giving rise to realistic masses 
and mixing patterns for the leptons and quarks, 
a unique feature of the  $SU(5)\times T'$ model 
is that the CP violation, predicted by the model, 
is entirely geometrical in origin. 
This interesting aspect of the model considered 
is a consequence of the special properties 
of the group $T'$. The model was shown  
to predict approximately 
tri-bimaximal neutrino (TBM) mixing with deviations 
which generate a non-zero $\sin\theta_{13} = 
\sin\theta_{c}/(3\sqrt{2})$, 
where $\theta_{c}$ is the Cabibbo angle, and 
a Dirac phase $\delta = 221^\circ \cong 5\pi/4$.
A rather detailed description of the lepton sector 
of the model is given in Section 2.

  We have derived first the predictions of the model 
for the light Majorana neutrino masses, $m_i$, $i=1,2,3$.
We have found that in the model considered
only light neutrino mass spectrum with 
{\it normal ordering} (or \emph{normal hierarchy}) 
is possible. In the standardly used convention 
this implies  $m_1 < m_2 < m_3$. 
The masses $m_i$ depend on two parameters of the model 
which, as we have shown, can be fixed using the values of 
$\Delta m^2_{21}$, which drives the solar neutrino 
oscillations, and the ratio 
$r = \Delta m^2_{21}/\Delta m^2_{31}$, 
$\Delta m^2_{31}$ being the neutrino mass squared 
difference responsible for the dominant 
atmospheric $\nu_{\mu}$ and $\bar{\nu}_{\mu}$ 
oscillations. Given the fact that 
$\Delta m^2_{21}$ and $r$ have been determined 
experimentally with rather small uncertainties, 
the light neutrino masses are predicted 
to lie in relatively narrow intervals 
around the values $m_1=1.14 \times 10^{-3}$ eV,  
$m_2=8.78 \times 10^{-3}$ eV,  $m_3=4.89 \times 10^{-2}$ eV 
(Fig. \ref{fig:spectrum}).
As a consequence, the model provides specific 
predictions for the sum of the light neutrino 
masses in terms of $\Delta m^{2}_{21}$ and $r$:
$m_1 + m_2 + m_3\cong \Delta m^{2}_{21}
(1 + (20r-1)^2 + (20r-3)^2)(1-(20r-1)^4)^{-\frac{1}{2}}$.
Numerically, we have found:
$m_1 + m_2 + m_3 \cong 5.9\times 10^{-2}~{\rm eV}$.

 The model provides also specific prediction for the 
effective Majorana mass in neutrinoless double beta
($\betabeta$-decay), $\meff$. The two Majorana phases 
$\alpha_{21}$ and $\alpha_{31}$ in the neutrino mixing
matrix (see eqs. (\ref{UPMNS} - (\ref{Q})) are predicted 
(to leading order) to have CP conserving values:
$\alpha_{21} = 0$ and $\alpha_{31} = \pi$. 
We found that the effective Majorana mass 
predicted by the model is given approximately by: 
$\meff \cong |(2m_1 + m_2)/3 - \sqrt{2}(m_2 - m_1)(\sin\theta_c)/9|$. 
Numerically, we obtained $\meff \cong 3.4\times 10^{-3}$ eV.

 In the last part of this work 
we have derived  detailed predictions 
for the rates of the lepton flavour violating (LFV) 
charged lepton radiative decays $\mu \rightarrow e + \gamma$, 
$\tau \rightarrow e + \gamma$ and $\tau \rightarrow \mu + \gamma$.
This was done within the commonly employed mSUGRA SUSY 
breaking scenario \cite{Chamseddine:1982jx}.
The values of the mSUGRA parameters $\tan\beta$ and $A_0$ 
are chosen from the intervals 
$\tan\beta= 3 \div 50$, $A_0= 0 \div 7m_0$, 
which are compatible with the constraints 
obtained by the ATLAS \cite{Aad:2011hh, Zhuang:2011qq} 
and CMS \cite{Khachatryan:2011tk} experiments 
at LHC. The values of the other two 
relevant mSUGRA parameters, $m_0$ and  $m_{1/2}$, 
are chosen from intervals favored by the global data 
analysis performed in \cite{Buchmueller:2011aa}. 
The data set used in this analysis includes in addition to 
the results of the ATLAS and the CMS  experiments, 
the data on the muon $(g-2)$, on the precision 
electroweak observables, on $B$-physics observables, 
astrophysical data on the cold dark matter density,
as well as the limits from the direct 
searches for Higgs boson and sparticles at LEP.
Based on the results obtained in 
\cite{Buchmueller:2011aa},  $m_0$ and  $m_{1/2}$
are chosen from, or to vary in, the intervals 
 $50~{\rm GeV} \leq m_0 \leq  400$ GeV and 
$300~{\rm GeV}\leq m_{1/2} \leq 800)$ GeV.
The branching ratios are 
calculated for masses of the heavy Majorana 
neutrinos $M_k = M = 10^{12}$ GeV. The  
GUT scale used is $M_X =2\times 10^{16}$ GeV. 
The results of the this analysis are reported graphically 
in Figs. \ref{fig:BRm0} - \ref{fig:Ratios}.

One specific prediction 
of the  $SU(5)\times T'$ model 
of flavour considered is that the quantity 
$|(Y_{\nu}Y^\dagger_{\nu})_{l l'}|$ 
on which the $l \rightarrow l^\prime + \gamma$ decay
branching ratios depend, $l=\mu,\tau$, 
$l'=e,\mu$ ($m_l > m_{l'}$), $Y_{\nu}$ 
being the matrix of neutrino Yukawa 
couplings in the basis in which the RH neutrino 
Majorana mass term and the matrix of charged lepton 
Yukawa couplings are diagonal, 
is a function of the PMNS neutrino 
mixing matrix $U$ and neutrino masses $m_i$ only,
$|(Y_{\nu}Y^\dagger_{\nu})_{l l^\prime}| =
|U_{l j}m_{j}U^*_{l^\prime j}|$.
Thus, in the model considered 
and for  given $l$ and $l^\prime$ 
this quantity is fixed numerically with small 
uncertainties. Together with the fact that the heavy 
Majorana neutrinos are degenerate in mass, 
this feature of the model implies that
the double ratios of the 
branching ratios 
$R(21/31)\equiv (BR(\mu\rightarrow e + \gamma)/
BR(\tau\rightarrow e + \gamma))
BR(\tau\rightarrow e \nu_{\tau} \bar\nu_e )$ and
$R(21/32)\equiv (BR(\mu\rightarrow e + \gamma)/
BR(\tau\rightarrow \mu + \gamma))
BR(\tau \rightarrow e\nu_{\tau} \bar\nu_e)$ 
do not depend not only on the mSUGRA parameters, 
but also on the heavy Majorana neutrino masses and the 
GUT scale $M_X$: 
$R(21/31) \cong |(Y_{\nu}Y^{\dagger}_{\nu})_{\mu e}|^2/
|(Y_{\nu}Y^{\dagger}_{\nu})_{\tau e}|^2$,
$R(21/32)\cong |(Y_{\nu}Y^{\dagger}_{\nu})_{\mu e}|^2/
|(Y_{\nu}Y^{\dagger}_{\nu})_{\tau \mu}|^2$. 
Numerically we have found:
$R(21/31) \cong 0.21$, $R(21/32) \cong 7.4\times 10^{-3}$.
These values are one of the characteristic predictions 
of the  model with $SU(5)\times T^\prime$ symmetry considered.
In particular, the $\tau \rightarrow \mu + \gamma$
decay branching ratio, $BR(\tau\rightarrow \mu + \gamma)/
BR(\tau \rightarrow e\nu_{\tau} \bar\nu_e )$, 
is predicted to be bigger than the
$\mu \rightarrow e + \gamma$ decay branching ratio,
$BR(\mu\rightarrow e + \gamma)$, by a factor of $1.36\times 10^{2}$.
We have found also that in a relatively 
large part of the  mSUGRA parameter space 
considered and for $M = 10^{12}$ GeV, 
the $\mu \rightarrow e \gamma$ 
decay branching ratio can satisfy the 
MEG upper bound $BR(\mu\rightarrow e + \gamma) < 2.4\times 10^{-12}$,
but still can have a value in the range of sensitivity 
of the MEG experiment, $BR(\mu\rightarrow e + \gamma) \gtap 10^{-13}$.

 The model with $SU(5)\times T^\prime$ symmetry proposed 
in ~\cite{Chen:2007afa,Chen:2009gf} and investigated 
in the present article provides a unified description
of the masses and mixing of the quarks and leptons, 
neutrinos included, as well as of 
the CP violation both in the quark and lepton sectors.
This makes it a viable model of flavour, which 
possesses a number of appealing features.
In the lepton sector the model provides 
specific predictions for 
the values of the light neutrino masses, the neutrino mass spectrum, 
the values of the neutrino mixing angles, 
including the smallest one $\theta_{13}$, and  the Dirac and Majorana 
CP violation phases in the neutrino mixing matrix.
All these predictions can and will be tested in the currently 
operating and future neutrino experiments. 
We are looking forward to the outcome of these tests.

\section*{Acknowledgments}
This work was supported in part by the INFN program 
on ``Astroparticle Physics'', by the Italian MIUR program 
on ``Neutrinos, Dark Matter and  Dark Energy in the Era of LHC''
(A.M. and S.T.P.) and by the World Premier International 
Research Center Initiative (WPI Initiative), 
MEXT, Japan  (S.T.P.).
The work of M-CC and KTM was supported in part, respectively
by the National Science Foundation 
under grant No. PHY-0970173 and
% The work of KTM was supported, in part, 
by the Department of Energy under 
Grant No. DE-FG02-04ER41290.

% \pagebreak
% \newpage


\begin{thebibliography}{99}
\bibitem{footn1}  This result can have important 
implications for the ``flavoured'' 
leptogenesis scenario of
generation of the baryon asymmetry of 
the Universe \cite{EMSTP08}.
\end{thebibliography}
\end{document}